\documentclass[
 preprint,superscriptaddress,showkeys,floatfix,amsmath,amssymb,prx,onecolumn,11pt,a4paper
]{revtex4-2}
\usepackage[utf8]{inputenc}
\usepackage{soul,xcolor}
\usepackage{graphicx}
\usepackage{hyperref}
\hypersetup{
    colorlinks=true,
    linkcolor=blue,
    citecolor=red,
    filecolor=magenta,      
    urlcolor=cyan}

\renewcommand{\selectlanguage}[1]{} 

\begin{document}

\title{In-operando control of sum-frequency generation in tip-enhanced nanocavities}

\author{Philippe Roelli}
\email{p.roelli@nanogune.eu}
\affiliation{%
  CIC nanoGUNE BRTA, 20018 Donostia-San Sebastián, Spain
}%

\author{Isabel Pascual}
\affiliation{%
  CIC nanoGUNE BRTA, 20018 Donostia-San Sebastián, Spain
}%
\affiliation{%
  Materials Physics Center, CSIC-UPV/EHU, 20018 Donostia-San Sebastián, Spain
}%

\author{Iris Niehues}
\affiliation{%
  Institute of Physics, University of Münster, 48149 Münster, Germany
  }%

\author{Javier Aizpurua}
\affiliation{%
  Donostia International Physics Center (DIPC), 20018 Donostia-San Sebastián, Spain
}%
\affiliation{%
  IKERBASQUE, Basque Foundation for Science, 48013 Bilbao, Spain
}%

\affiliation{%
Department of Electricity and Electronics, University of the Basque Country (UPV/EHU), 48940 Leioa, Spain
}%

\author{Rainer Hillenbrand}
\email{r.hillenbrand@nanogune.eu}
\affiliation{%
  CIC nanoGUNE BRTA, 20018 Donostia-San Sebastián, Spain
}%
\affiliation{%
  IKERBASQUE, Basque Foundation for Science, 48013 Bilbao, Spain
}%
\affiliation{%
Department of Electricity and Electronics, University of the Basque Country (UPV/EHU), 48940 Leioa, Spain
}%

\date{\today}

\begin{abstract}
Sum-frequency generation (SFG) is a second-order nonlinear process widely used for characterizing surfaces and interfaces with monolayer sensitivity. 
Recently, optical field enhancement in plasmonic nanocavities has enabled SFG with continuous wave (CW) lasers from nanoscale areas of molecules, 
promising applications like nanoscale SFG spectroscopy and coherent upconversion for mid-infrared detection at visible frequencies. 
Here, we demonstrate CW SFG from individual nanoparticle-on-mirror (NPoM) cavities, which are resonant at visible frequencies and filled with a monolayer of molecules, 
when placed beneath a metal scanning probe tip. 
The tip acts as an efficient broadband antenna, focusing incident CW infrared illumination onto the nanocavity. 
The cascaded near-field enhancement within the NPoM nanocavity 
yields nonlinear optical responses across a broad range of infrared frequencies, 
achieving SFG enhancements of up to 14 orders of magnitude.   
Further, nanomechanical positioning of the tip allows for in-operando control of SFG by tuning the local field enhancement rather than the illumination intensities. 
The versatility of tip-enhanced nanocavities allows for SFG studies of a wide range of molecular species in the few-molecule regime 
without the need for complex nanofabrication of doubly-resonant nanocavities.
Our results also promise SFG nanoimaging with tips providing strong visible and IR field enhancement at their apex, 
offering a robust platform for future applications in nonlinear nanooptics.
\end{abstract}

\maketitle

\section{Introduction}

Sum-frequency generation (SFG) is a coherent second-order nonlinear process 
\cite{dick_spectroscopy_1983,zhu_surface_1987,shen_fundamentals_2016}
with significant applications in up-conversion-based and vibration-selective imaging and spectroscopy \cite{fischer_nonlinear_2005,chung_biomolecular_2013,shah_chemical_2020,wang_vibrational_2021}. 
The SFG process is particularly effective for probing vibrational modes at interfaces 
where the material second-order ($\chi^{(2)}$) nonlinearity is often activated.  
By carefully tuning the infrared (IR) frequency to the vibrational mode $\nu$ and phase matching IR and visible (VIS) input beams, 
a coherent SFG output beam at frequency 
$\omega_{\mathrm{SFG}}=\omega_{\mathrm{IR}}(\nu)+\omega_{\mathrm{VIS}}$ can be detected, revealing subtle details about the molecular environment and bonding at the interface. 

Compared to coherent anti-stokes Raman spectroscopy (CARS), vibrational spectroscopy via SFG offers the advantage of requiring only a single Raman process, 
the second Raman process being replaced by a resonant IR process. 
This results in SFG having significantly higher intrinsic cross-sections than CARS. 
However, both techniques suffer from diffraction limitations, hindering nanoscale spatial resolution and the observation of few or even single molecules. 
CARS overcomes this limitation using plasmonic nanoparticles or scanning probe tips, which confine and enhance fields, 
enabling enhanced sensitivity and spatial resolution for sensing and imaging \cite{chen_surface_1979,ichimura_tip-enhanced_2004}. 
Additionally, surface- and tip-enhanced CARS enables time-resolved monitoring of vibrations reaching the single molecule level \cite{yampolsky_seeing_2014,luo_imaging_2023}. 

Surprisingly, the combination of SFG with optical cavities remains a widely unexplored terrain. 
Only recently, cavity-enhanced SFG has been implemented \cite{chen_continuous-wave_2021} 
in self-assemblies of organic molecules with the help of nanoparticle-on-mirror (NPoM) structures \cite{baumberg_extreme_2019}, 
in which the incident VIS and IR fields are concentrated in a 1 nm high and a few 10 nm wide gap 
between a particle and a mirror, defined by the thickness of the molecular layer and the facet of the particle, respectively. 
When the VIS-resonant NPoM cavity is combined with an IR-resonant antenna in a carefully engineered way, 
the infrared field enhancement in the NPoM gap can be strongly boosted. 
In this case, remarkable SFG and difference frequency generation (DFG) signals 
can be observed simultaneously even with low power continuous wave (CW) illumination, 
opening a new path for efficient up-conversion and detection of IR photons with visible cameras \cite{roelli_molecular_2020}. 
  
So far, efficient cavity-enhanced SFG relies on sophisticated fabrication of doubly resonant cavities 
and may require design variation for the study of different molecular vibrational modes. 
Importantly, the spatial overlap of IR and VIS fields inside the gap has to be well-defined 
to maximize signal yield and to avoid spurious cavity responses 
that may prevent the observation of clear SFG signals \cite{xomalis_detecting_2021}. 
Further, because of the relatively inefficient coupling between far-field radiation and plasmonic nanocavities, 
the enhanced SFG signal fails to approach the up-conversion efficiencies of optomechanical devices 
operating in different frequency ranges \cite{chu_perspective_2020}. 
For all these reasons, a more versatile and active platform 
allowing for the adjustment of spectral and spatial characteristics of the cavity is desirable, 
for example, to push SFG spectroscopy towards single molecule sensing and 
for evaluating a large number of molecular species for future IR to VIS up-conversion applications. 

Here we demonstrate efficient CW infrared to visible SFG and DFG in NPoM nanocavities filled with a monolayer of organic molecules, 
which is enhanced and controlled in-operando with the metal tip of a scattering-type scanning near-field optical microscope (s-SNOM). 
The tip acts as an antenna in both IR and VIS spectral ranges, concentrating the incident fields of both frequencies at its apex. 
The apex fields serve as a local illumination for the NPoM cavity, enhancing the fields in the gap between the particle and mirror. 
By recording second-order nonlinear responses (SFG and DFG) 
under infrared illumination at two distinct frequencies, 
separated by $500$~cm$^{-1}$ 
and both tuned to specific molecular vibrational modes, 
we demonstrate efficient coherent upconversion across the whole mid-infrared spectral range. 
Key to this achievement is the non-resonant but strong IR field concentration at the tip apex of the s-SNOM.  
Numerical simulations of the relevant field enhancements verify our experimental results 
and let us explain the remarkable upconversion signals through a cascaded near-field enhancement, similar 
to that of an efficient nanolens made of a chain of metal nanoantennas \cite{li_self-similar_2003,hoppener_self-similar_2012}, 
and through momentum conservation facilitated by the broad momentum distribution of the near fields \cite{novotny_principles_2012} in the NPoM nanocavities.
Our study not only highlights cascaded enhancements of the nonlinear response of molecular vibrations up to 14 orders of magnitude 
but also introduces a method for actively controlling field enhancements at visible and infrared frequencies 
through precise 3D nanomechanical positioning of the tip. 

\newpage

\section{Results}

\begin{figure}[ht!]
\centering
\includegraphics[scale=0.6]{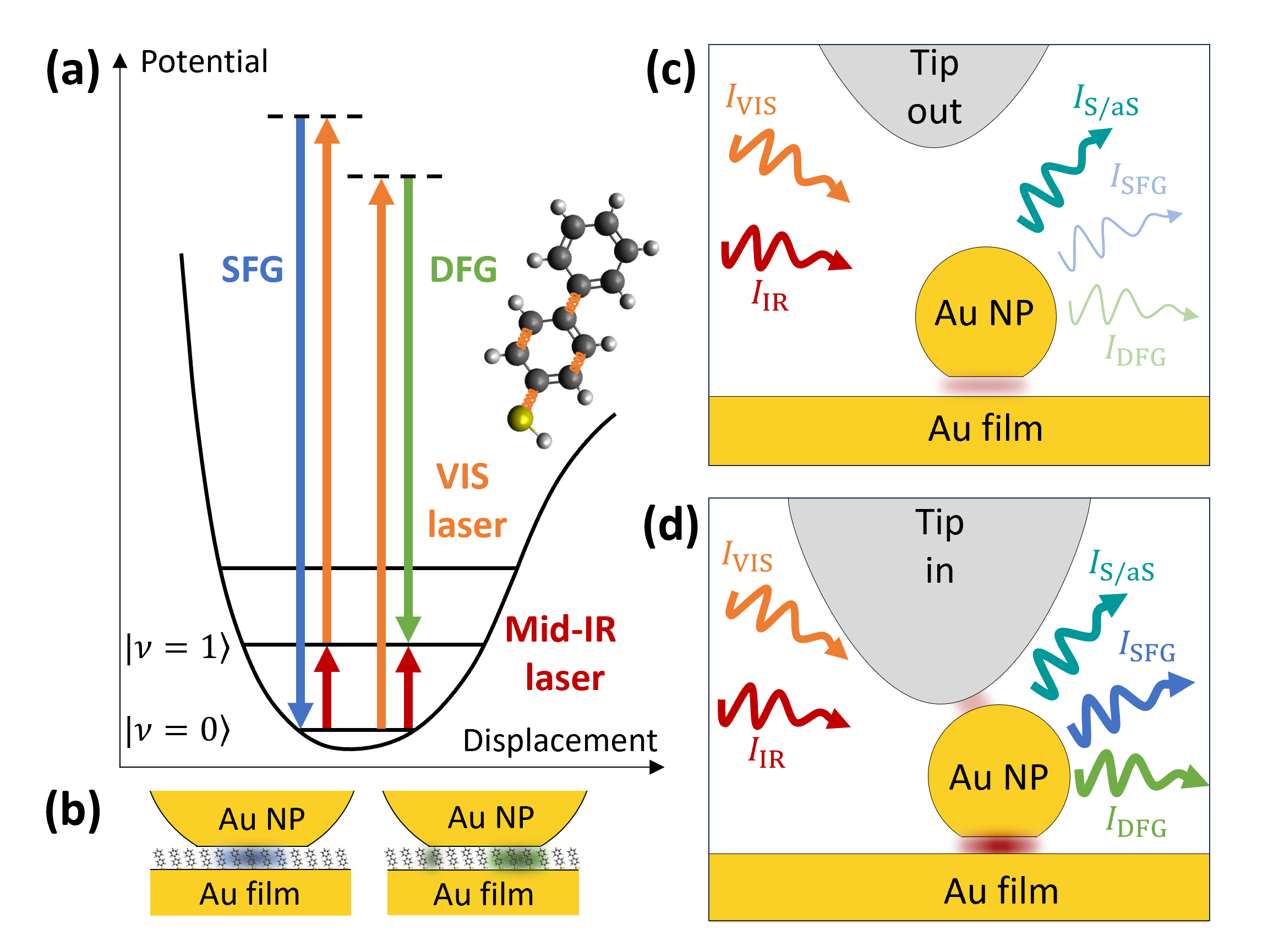}
  \caption{
  \textbf{Nonlinear optics inside a tip-enhanced NPoM cavity}.
  (a) 
  Jablonski energy diagram of vibrational coherent upconversion signals for one mode of the BPT molecule (inset). 
  (b) Illustration of the field distributions inside the BPT functionalized NPoM gap at the frequencies where SFG (blue) and DFG (green) occur. 
  (c,d) Illustration of vibrational scattering from a tip-controlled NPoM cavity in two configurations: 
  tip out of contact (c), tip in contact with the NPoM (d). 
  The molecules in the NPoM cavity, under both infrared ($I_{\mathrm{IR}}$) and visible ($I_{\mathrm{VIS}}$) illuminations, 
  generate Stokes/anti-Stokes (S/aS), SFG and DFG vibrational signals. 
  }
  \label{fig:sketch-conversion}
\end{figure}

The concept of tip-controlled nonlinear optics based on a molecules-filled NPoM cavity is outlined in Fig.~\ref{fig:sketch-conversion}. 
We first illustrate in Fig.~\ref{fig:sketch-conversion}a the intrinsic vibrational SFG and DFG processes of an organic molecule using a Jablonski energy diagram. 
When molecules are simultaneously illuminated by visible light (orange arrow) and by a IR radiation (red arrow) 
tuned to a molecular vibrational mode both Raman and IR active, i.e. $\chi^{(2)}$ active \cite{shen_fundamentals_2016}, 
the targeted vibrational mode mediates the generation of SFG and DFG signals 
at the angular frequencies $\omega_{+}=\omega_{\mathrm{VIS}}+\omega_{\mathrm{IR}}$ (upper vibrational sideband) 
and $\omega_{-}=\omega_{\mathrm{VIS}}-\omega_{\mathrm{IR}}$ (lower vibrational sideband) respectively. 
During these nonlinear optical processes, the vibrational mode is initially brought to its first excited state 
by resonant IR photons (red arrow in in Fig.~\ref{fig:sketch-conversion}a) 
and then transitions via Raman scattering to its ground state (SFG, blue arrow) or to its second excited state (DFG, green arrow). 

To enhance vibrational optical processes, we embed organic molecules into NPoM cavities, that is, into the gap between a gold nanoparticle and a gold mirror 
(illustrated in  Fig.~\ref{fig:sketch-conversion}b). 
Under VIS illumination alone, the Stokes (S) and anti-Stokes (aS) signals intensities, $I_{\mathrm{S/aS}}$ (cyan arrow in Fig.~\ref{fig:sketch-conversion}c,d), 
are enhanced by factors $\mathcal{F}_{\mathrm{S}}$ and $\mathcal{F}_{\mathrm{aS}}$, which are given by \cite{ru_principles_2008}:  
\begin{equation}
\mathcal{F}_{\mathrm{S/aS}}=F_{\mathrm{VIS}}\:F_{\mathrm{-/+}},   
\label{eq:f-sers}
\end{equation}
where $F_{\mathrm{VIS}}$, $F_{\mathrm{+}}$ and $F_{\mathrm{-}}$ are the position-dependent near-field intensity enhancement factors 
inside the NPoM gap at the angular frequencies $\omega_{\mathrm{VIS}}$, $\omega_{+}$ and $\omega_{-}$ (surface-enhanced Raman scattering, SERS). 
Under additional IR illumination, the signal enhancement factors of vibrational SFG and DFG 
(light blue and green arrows) for a molecule at a specific location in the gap can be described by \cite{chen_continuous-wave_2021}: 
\begin{align}
 \mathcal{F}_{\mathrm{SFG}}& = F_{\mathrm{IR}}\:F_{\mathrm{VIS}}\:F_{\mathrm{+}}
 \label{eq:f-sfg} \\
 \mathcal{F}_{\mathrm{DFG}}& = F_{\mathrm{IR}}\:F_{\mathrm{VIS}}\:F_{\mathrm{-}},  
 \label{eq:f-dfg}
\end{align}
where $F_{\mathrm{IR}}$ is the position-dependent near-field intensity enhancement factor inside the NPoM gap at the angular frequency $\omega_{\mathrm{IR}}$. 
Due to the absence of both plasmonic and geometric resonances in small Au particles at IR frequencies, $F_{\mathrm{IR}}$ remains weak. 
Consequently, SFG and DFG signals of intensity $I_{\mathrm{SFG/DFG}}\propto \int_{\mathrm{V_{\mathrm{mol}}}} \mathcal{F}_{\mathrm{SFG/DFG}}\mathrm{dV}$ 
\cite{talley_surface-enhanced_2005}, where $\mathrm{V_{\mathrm{mol}}}$ indicates the volume occupied by molecules, 
are too small to be detected with the use of an NPoM cavity  
\footnote{We omit potential phase and collective vibrational effects, as their contributions within a cavity are still debated \cite{zhang_optomechanical_2020,heeg_experimental_2021}.}. 
To circumvent this problem, we place a nanomechanically controlled metal tip above the NPoM cavity (Fig.~\ref{fig:sketch-conversion}b,c). 
The tip primarily serves as a broadband non-resonant infrared antenna, concentrating the incident IR field at its apex. 
In close proximity to the Au nanoparticle (Fig.~\ref{fig:sketch-conversion}c),  
the near field at the tip apex provides an additional illumination of the NPoM cavity, 
increasing substantially $F_{\mathrm{IR}}$ 
and thus the SFG and DFG signal intensities.  
We note that under visible and infrared CW illuminations both anti-Stokes ($I_{\mathrm{aS}}$, linear) and SFG ($I_{\mathrm{SFG}}$, nonlinear) signals 
appear simultaneously at $\omega_{+}$, whereas $I_{\mathrm{S}}$ and $I_{\mathrm{DFG}}$ overlap at $\omega_{-}$. 
In general, the signals appearing on these upper and lower vibrational sidebands, 
$I_{\mathrm{+}}=I_{\mathrm{aS}}+I_{\mathrm{SFG}}=c_{\mathrm{aS}}\:\mathcal{F}_{\mathrm{aS}}+c_{\mathrm{SFG}}\:\mathcal{F}_{\mathrm{SFG}}$ 
and $I_{\mathrm{-}}=I_{\mathrm{S}}+I_{\mathrm{DFG}}=c_{\mathrm{S}}\:\mathcal{F}_{\mathrm{S}}+c_{\mathrm{DFG}}\:\mathcal{F}_{\mathrm{DFG}}$, 
respectively, depend on molecular and gold properties at the interfaces of the NPoM cavity.  
These properties are described by the coefficients 
$c_{\,\mathrm{aS}}$, $c_{\,\mathrm{SFG}}$, $c_{\,\mathrm{S}}$ and $c_{\,\mathrm{DFG}}$ 
, which are challenging to evaluate. 
This challenge includes quantifying the non-resonant SFG signal from the gold at the interfaces 
\cite{covert_assessing_2015,dalstein_nonlinear_2018}, 
which adds coherently to the resonant vibrational SFG signal from the molecular layer 
\cite{humbert_sum-frequency_2019}. 
We thus only calculate intensity enhancement factors in this study, 
as they fully capture the role of the tip on both the linear and nonlinear optical scattering processes. 

\newpage

\begin{figure}[ht!]
\centering
\includegraphics[scale=0.60]{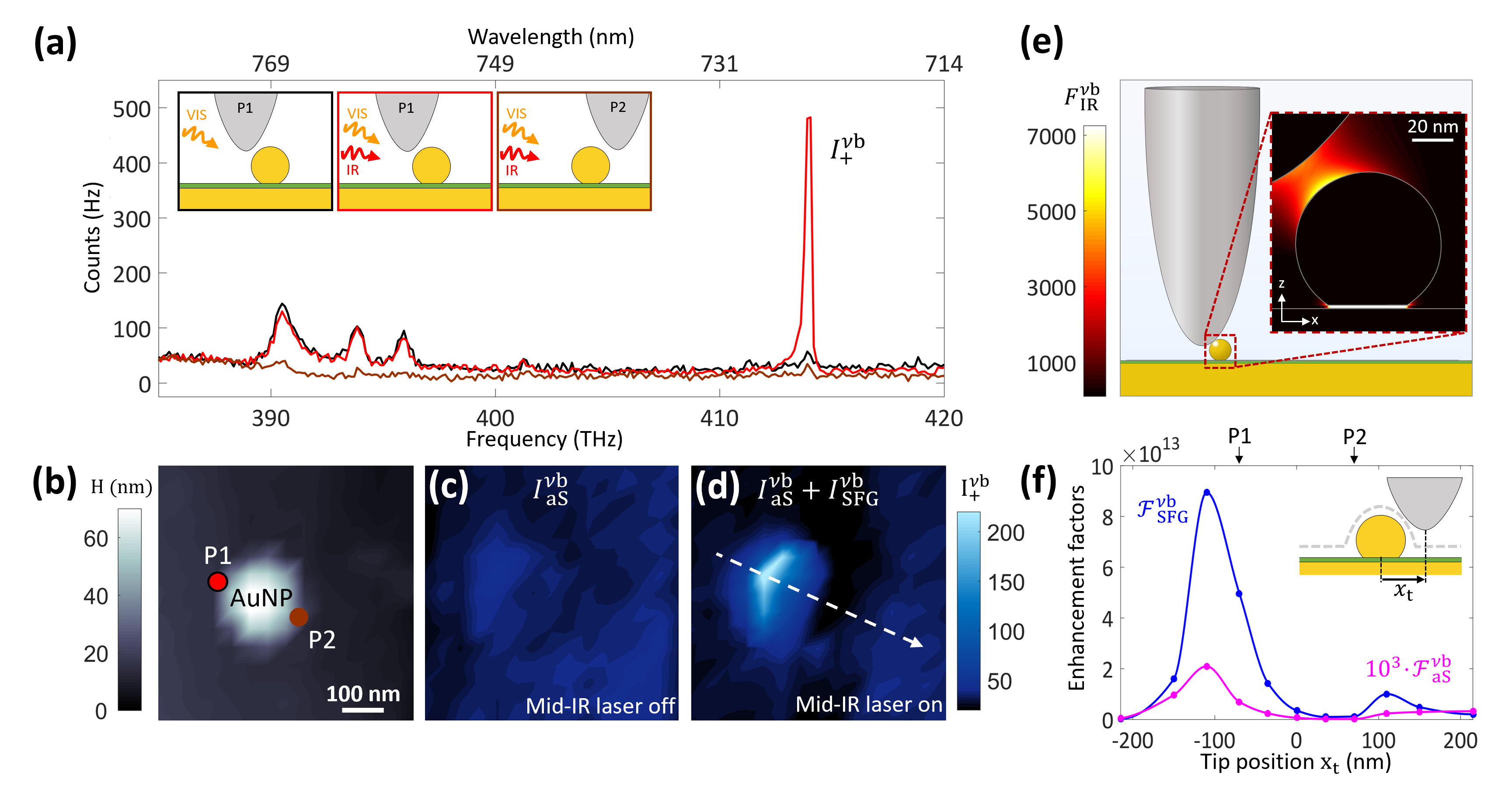}
  \caption{\textbf{Tip-controlled cavity-enhanced SFG}. 
  (a) Spectra of a BPT-filled NPoM cavity below an oscillating metallized scanning probe tip. 
  The NPoM is formed by a 63~nm high faceted Au particle on top of an Au mirror that is functionalized with a monolayer of BPT molecules. 
  Insets illustrate the positions P1 and P2 of the tip relative to particle, as well as illumination at VIS (orange) and IR (red) frequencies. 
  Black spectrum is recorded for tip positioned at P1 under VIS illumination exclusively. 
  Red and brown spectra were recorded under VIS and IR 
  ($2.2$~mW tuned to the vibrational mode $\nu_{\mathrm{b}}$ at 32~THz) 
  illumination at positions P1 and P2, respectively.  
  (b) Topography image of the NPoM cavity. Red and brown dots indicate the tip positions P1 and P2 where spectra shown in panel (a) were recorded. 
  Maps of $I^{\nu_{\mathrm{b}}}_+$ (averaged over peak area) without (c) and with (d) IR illumination. 
  VIS illumination is at $382$~THz ($785$~nm) at a power of $200$~$\mu$W. 
  Acquisition time per spectra is $2$~s. Tapping amplitude (TA) is $50$~nm. 
  (e) Simulated spatial distribution of the intensity enhancement factor at 32~THz $F_{\mathrm{IR}}^{\nu_{\mathrm{b}}}$ 
  around the NPoM cavity. 
  (f) Simulated $\mathcal{F}_{\mathrm{aS}}^{\nu_{\mathrm{b}}}(\vec{\mathrm{r}}_{\mathrm{hs}})$ (pink line) 
  and $\mathcal{F}_{\mathrm{SFG}}^{\nu_{\mathrm{b}}}(\vec{\mathrm{r}}_{\mathrm{hs}})$ (blue line) enhancement factors along the dashed arrow depicted in (d). 
  $x_{\mathrm{t}}$ is the lateral position of the tip with respect to the particle's center.
  The vertical distance $\Delta z_{\mathrm{t}}=20$~nm between tip apex and NPoM is kept constant, as illustrated in the inset. 
  Optical maps for other vibrational signals can be found in Supplementary Note~3. 
  }
  \label{fig:hs-sfg}
\end{figure}


In Fig.~\ref{fig:hs-sfg}a, we experimentally demonstrate that 
(i) the tip enables detection of otherwise imperceptible SFG signals from a molecule-filled NPoM cavity under VIS and IR continuous wave illumination and 
(ii) this signal strongly depends on the mechanical nanopositionning of the tip. 
To that end, we fabricate NPoMs by functionalizing a template-stripped gold film with a monolayer of biphenyl-4-thiol (BPT) molecules 
and drop-casting faceted gold particles. 
Thus, the NPoM gap is fully filled with molecules and the gap height is determined by the monolayer thickness \cite{ahmed_structural_2021} (see Methods). 
We probe the NPoM cavities with an s-SNOM, utilizing an oscillating metallized atomic force microscopy (AFM) tip (nano-FTIR tip with nominal apex diameter of 100~nm). 
In s-SNOM, an off-axis parabolic mirror focuses a laser beam from the side onto the tip and collects the elastically back-scattered light, 
enabling operation with multiple collinear beams across a wide THz to VIS frequency range. 
With a grating spectrometer, the setup also detects inelastically scattered light, including Raman signals \cite{kusch_dual-scattering_2017}, 
making it ideal for our studies. More details about the setup and alignment can be found in the Methods and Supplementary Note~1. 

We first place the tip above the NPoM at position $\mathrm{P1}$ (illustrated in the inset of Fig.~\ref{fig:hs-sfg}a 
and indicated in the topography image of Fig.~\ref{fig:hs-sfg}b), illuminate the tip above the NPoM cavity with a focused CW VIS laser beam of $382$~THz ($785$~nm) 
and record a spectrum of the inelastically scattered light in the absence of IR illumination (black spectrum, in Fig.~\ref{fig:hs-sfg}a). 
This spectrum shows the anti-Stokes signals corresponding to different vibrational modes of BPT molecules inside the NPoM cavity.  
Interestingly, when a CW monochromatic IR illumination is tuned to the molecule's bending mode, labelled $\nu_{\mathrm{b}}$, at 
$\omega_{\mathrm{IR}}^{\nu_{\mathrm{b}}}/(2\pi)=32$~THz ($1080$~cm$^{-1}$), 
we observe a remarkable and frequency-selective increase of the intensity on the sideband $\omega^{\nu_{\mathrm{b}}}_+$ 
($I^{\nu_{\mathrm{b}}}_{+}\equiv I_{+}(\nu_{\mathrm{b}})$, red spectrum). 
In agreement with a previous far-field spectroscopy study \cite{chen_continuous-wave_2021}, 
the increased peak intensity may result from an enhancement of the resonant SFG signal associated with the vibrational mode $\nu_{\mathrm{b}}$ or, 
in other words, to a coherent upconversion of IR photons from $32$~THz to $414$~THz mediated by the molecular vibration $\nu_{\mathrm{b}}$. 
In the future, frequency-scan SFG experiments could be conducted to quantify the vibrational SFG contribution.
We note that the peak intensities of the other vibrational sidebands, which are not driven by the CW monochromatic IR illumination, remain unchanged. 
This observation rules out the possibility that $I^{\nu_{\mathrm{b}}}_{+}$ is enhanced due to heating effects caused by the IR illumination.

To explore the impact of the tip position on the SFG signals, we record spectra while scanning laterally the tip across the NPoM cavity. 
At position $\mathrm{P2}$ (brown curve in Fig.~\ref{fig:hs-sfg}a), all peaks show a reduced intensity compared to position $\mathrm{P1}$. 
Mapping the anti-Stokes and SFG signals as function of tip position (Fig.~\ref{fig:hs-sfg}c,d) 
reveals strong SFG signals for tip positions around $\mathrm{P1}$, 
whereas anti-Stokes signals remain weak regardless of the tip position. 
These measurements highlight the dominance of tip-enhanced SFG over anti-Stokes signals 
and evidence an asymmetry of tip-enhanced SFG with respect to the particle center. 
This asymmetry is a robust feature observed for many particles we imaged. Few examples are shown in Supplementary Note~4.
 
The asymmetry of the SFG map of Fig.~\ref{fig:hs-sfg}d can be verified by 
numerical calculations of the signal enhancement factors $\mathcal{F}_{\mathrm{SFG}}^{\nu_{\mathrm{b}}}$ and $\mathcal{F}_{\mathrm{aS}}^{\nu_{\mathrm{b}}}$ 
as a function of the lateral position $x_{\mathrm{t}}$ of the tip with respect to the particle's center. 
To that end, the tip is modelled as a $1~\mu$m long Pt half ellipsoid with an apex diameter of 100~nm and the NPoM 
as a truncated 60~nm height Au sphere at a distance of 1.15~nm above an Au mirror (see schematic of Fig.~\ref{fig:hs-sfg}e and Methods). 
For a given tip position, we calculate the spatial distributions 
$F_{\mathrm{IR}}^{\nu_{\mathrm{b}}}(\Vec{r})$, $F_{\mathrm{VIS}}(\Vec{r})$ and $F_{+}^{\nu_{\mathrm{b}}}(\Vec{r})$ around the tip and NPoM cavity.
In the inset of Fig.~\ref{fig:hs-sfg}e we show exemplarily $F_{\mathrm{IR}}^{\nu_{\mathrm{b}}}(\Vec{r})$ for tip at position $\mathrm{P1}$. 
Multiplying the intensity distributions, we extract the maximal enhancement factors 
for the SFG and aS signals inside the gap of the NPoM cavity (according to Eqs.~\ref{eq:f-sers}~and~\ref{eq:f-sfg}, see also Supplementary Note~5 and~6). 
We refer to the position at which the factors are maximal as hot-spots ($\vec{\mathrm{r}}_{\mathrm{hs}}$) and write the corresponding signal enhancement factors as $\mathcal{F}_{\mathrm{SFG}}^{\nu_{\mathrm{b}}}(\vec{\mathrm{r}}_{\mathrm{hs}})$ and $\mathcal{F}_{\mathrm{aS}}^{\nu_{\mathrm{b}}}(\vec{\mathrm{r}}_{\mathrm{hs}})$. 
The blue curve in Fig.~\ref{fig:hs-sfg}f shows $\mathcal{F}_{\mathrm{SFG}}^{\nu_{\mathrm{b}}}(\vec{\mathrm{r}}_{\mathrm{hs}})$ when the tip is scanned across the particle, as illustrated in the inset. 
We find that $\mathcal{F}_{\mathrm{SFG}}^{\nu_{\mathrm{b}}}(\vec{\mathrm{r}}_{\mathrm{hs}})$ is asymmetric with respect to the particle, 
confirming the experimental SFG results (Fig.~\ref{fig:hs-sfg}d). 
Further, we find that the SFG signal enhancement can reach up to 14 orders of magnitude for a small range of tip positions close to $\mathrm{P1}$.  
For comparison, the pink curve in Fig.~\ref{fig:hs-sfg}f depicts the simulated linescan of the anti-Stokes signal enhancement, $\mathcal{F}_{\mathrm{aS}}^{\nu_{\mathrm{b}}}(\vec{\mathrm{r}}_{\mathrm{hs}})$ at the frequency of the vibrational mode $\nu_{\mathrm{b}}$. 
$\mathcal{F}_{\mathrm{SFG}}^{\nu_{\mathrm{b}}}(\vec{\mathrm{r}}_{\mathrm{hs}})$ is about 3 to 4 orders of magnitude increased 
as compared to $\mathcal{F}_{\mathrm{aS}}^{\nu_{\mathrm{b}}}(\vec{\mathrm{r}}_{\mathrm{hs}})$
, because of the infrared intensity enhancement 
provided by the tip-enhanced NPoM cavity. 
Interestingly, $\mathcal{F}_{\mathrm{aS}}^{\nu_{\mathrm{b}}}(\vec{\mathrm{r}}_{\mathrm{hs}})$ qualitatively follows 
the $\mathcal{F}_{\mathrm{SFG}}^{\nu_{\mathrm{b}}}(\vec{\mathrm{r}}_{\mathrm{hs}})$ linescan, that is, it exhibits a maximum near tip position $\mathrm{P1}$ 
and a similar asymmetry with respect to the particle (alike results of Fig.~\ref{fig:hs-sfg}c). 
This maximum is 17 times higher than the anti-Stokes signal enhancement of the NPoM cavity alone, 
indicating that the tip also enhances near fields in the NPoM gap at visible frequencies. 
We note that the asymmetry of both SFG and aS linescans can be attributed to the combined and enhanced 
longitudinal and transversal response of the tip under visible illumination (see Supplementary Note~8).   

\newpage

\begin{figure}[ht!]
\centering
\includegraphics[scale=0.58]{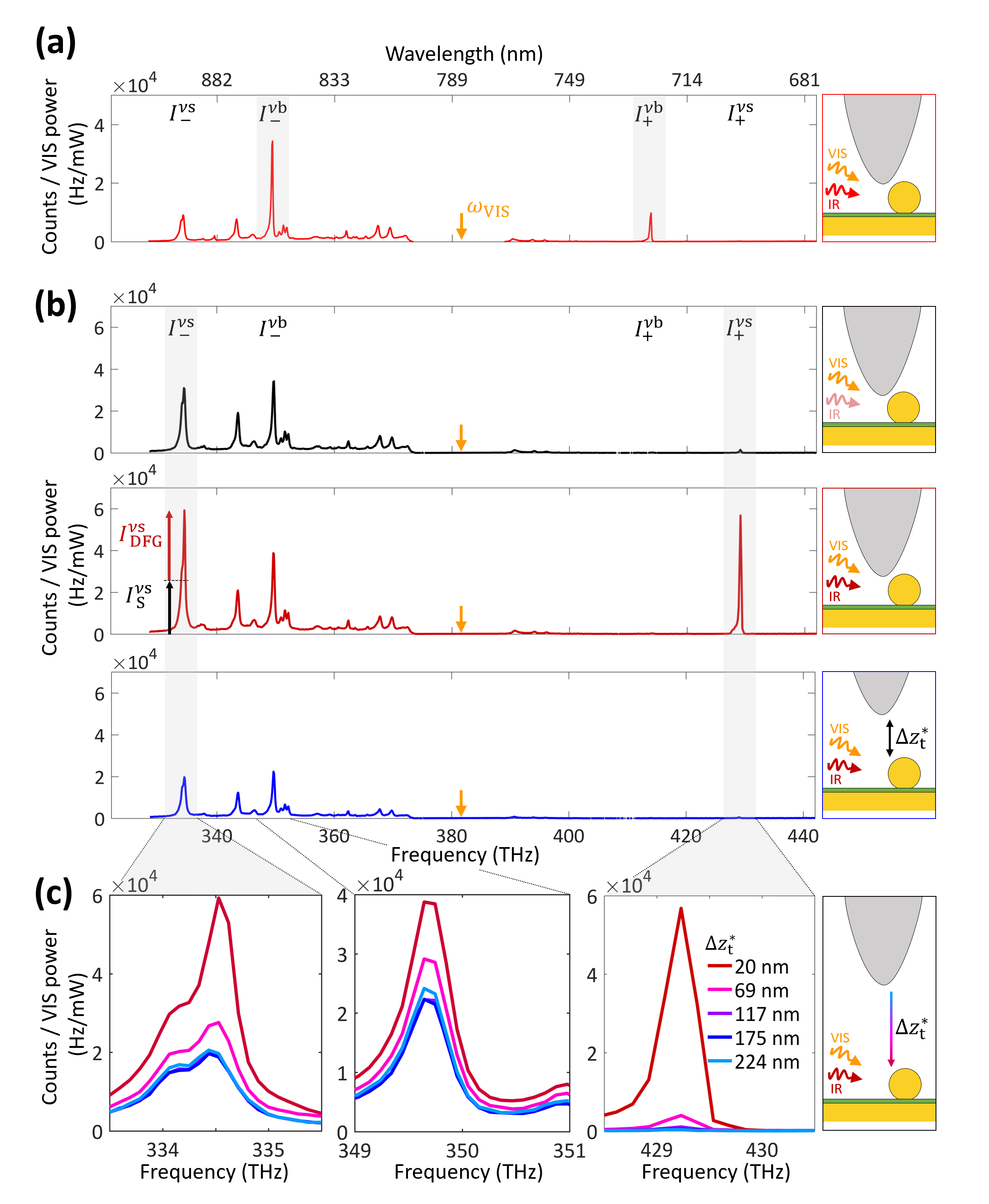}
  \caption{ \textbf{Tip-height dependent vibrational spectra}.  
  (a) Spectrum of a BPT-filled NPoM cavity below an oscillating metallized scanning probe tip in contact at position $\mathrm{P1}$
  under VIS 
  ($200$~$\mu$W at $382$~THz / 785~nm, marked $\omega_{\mathrm{VIS}}$)
  and intense IR ($6.2$~mW tuned to the vibrational mode $\nu_{\mathrm{b}}$ at 32~THz / 1080 ~cm$^{-1}$) illumination. 
  Data are the same as the one shown in Fig.~\ref{fig:hs-sfg}a.
  (b) Spectra of BPT molecules in another NPoM cavity below an oscillating metallized scanning probe tip. 
  Black spectrum was recorded for tip in contact ($\Delta z_{\mathrm{t}}^*=20$~nm) 
  under VIS ($250$~$\mu$W at $382$~THz) and weak IR ($0.2$~mW tuned to the vibrational mode $\nu_{\mathrm{s}}$ at 48~THz / 1585~cm$^{-1}$) illumination. 
  Red and blue spectra were recorded under VIS and intense IR 
  ($5.8$~mW) illumination in contact ($\Delta z_{\mathrm{t}}^*=20$~nm) and in retracted ($\Delta z_{\mathrm{t}}^*=175$~nm) positions, respectively.
  (c) Tip-height variation of signals' intensities around specific vibrational sidebands for $5.8$~mW IR illumination. 
  From left to right, we show 
  $I^{\nu\mathrm{s}}_-$, $I^{\nu_{\mathrm{b}}}_-$ and $I^{\nu\mathrm{s}}_+$. 
  Sketches illustrate the tip at position $\mathrm{P1}$ for two tip-NPoM effective distances $\Delta z_{\mathrm{t}}^*$ (in contact and retracted) 
  as well as illumination at VIS (orange) and IR (red) frequencies.
  Acquisition time for spectra is $2$~s. Tapping amplitude (TA) is $50$~nm. 
  }
  \label{fig:control-conversion}
\end{figure}

Tip-enhanced nano-cavities can be applied to study nonlinear CW optical signals 
from nanoscale areas of molecular monolayers across a broad infrared spectral range.
This is illustrated in Fig.~\ref{fig:control-conversion}, 
where two spectra recorded at tip position $\mathrm{P1}$ are compared. 
In Fig.~\ref{fig:control-conversion}a, we show the spectrum obtained under illumination 
at $\omega_{\mathrm{IR}}^{\nu_{\mathrm{b}}}/(2\pi)=32$~THz (same as in Fig.~\ref{fig:hs-sfg}a). 
In Fig.~\ref{fig:control-conversion}b, a different vibrational mode of the BPT molecules 
is driven by IR illumination from a quantum cascade laser (QCL). 
For the latter, we selected the molecular stretching mode (labelled $\nu_{\mathrm{s}}$) 
at $\omega_{\mathrm{IR}}^{\nu_{\mathrm{s}}}/(2\pi)=48$~THz.  
The red spectrum in Fig.~\ref{fig:control-conversion}b shows that the intensity 
at the sideband $\omega^{\nu_{\mathrm{s}}}_+$ is enhanced by a factor of 36 
compared to the black spectrum in Fig.~\ref{fig:control-conversion}b 
when the QCL illumination power is increased by a factor of 32. 
The two sidebands driven in Fig.~\ref{fig:control-conversion}a and Fig.~\ref{fig:control-conversion}b differ by 16~THz (500~cm$^{-1}$), 
indicating that the tip efficiently enhances IR radiation and thus SFG signals over a broad spectral range.
Interestingly, IR illumination at $\omega_{\mathrm{IR}}^{\nu_{\mathrm{s}}}$ 
also increases the intensity of the $\omega^{\nu_{\mathrm{s}}}_-$ sideband (Fig.~\ref{fig:control-conversion}b), 
while illumination at $\omega_{\mathrm{IR}}^{\nu_{\mathrm{b}}}$ increases the intensity 
of the $\omega^{\nu_{\mathrm{b}}}_-$ sideband (Fig.~\ref{fig:control-conversion}a). 
This demonstrates frequency-selective tip-enhanced DFG alongside mode-selective tip-enhanced SFG.

Employing a metallic scanning probe tip, we can control in-operando 
the intensity enhancement factor inside the NPoM gap. 
For an experimental demonstration, we record spectra while retracting the tip. 
Fig.~\ref{fig:control-conversion}c shows close-ups of vibrational signals around 
the sidebands $\omega^{\nu_{\mathrm{b}}}_-$ (only Stokes signal), 
$\omega^{\nu_{\mathrm{s}}}_-$ (DFG and Stokes signals) 
and $\omega^{\nu_{\mathrm{s}}}_+$ (comprising SFG and a marginal anti-Stokes contribution) 
for various effective vertical tip-NPoM distances ($\Delta z_{\mathrm{t}}^*$, see Methods). 
We observe that the SFG signal is efficiently controlled over several orders of magnitude, 
vanishing to the noise limit when the tip is retracted by more than 250~nm. 
Further, the driven $I^{\nu_{\mathrm{s}}}_-$ and non-driven $I^{\nu\mathrm{b}}_-$ 
signals can be controlled too. 
Although they decrease with increasing $\Delta z_{\mathrm{t}}^*$, 
they do not vanish completely. 
We attribute the residual signals to those originating from the NPoM cavity in absence of tip
and refer to them as background signals $I_{\mathrm{-,\: bg}}$. 
Overall, tip nanopositioning allows for increasing nonlinear optical signals 
via controlling intensity enhancement factors rather than 
by augmenting the illumination power. 
Tip-enhanced NPoMs offer therefore a means for efficient IR to VIS coherent upconversion, 
not only preventing heating of delicate NPoM cavities \cite{ahmed_structural_2021} 
, but also offering unprecedented opportunities 
for active tuning of the coupling regime 
\cite{park_tip-enhanced_2019,darlington_highly_2023} 
between nanocavity and  molecular vibration \cite{metzger_purcell-enhanced_2019}. 

\newpage

\begin{figure}[ht!]
\centering
\includegraphics[scale=0.65]{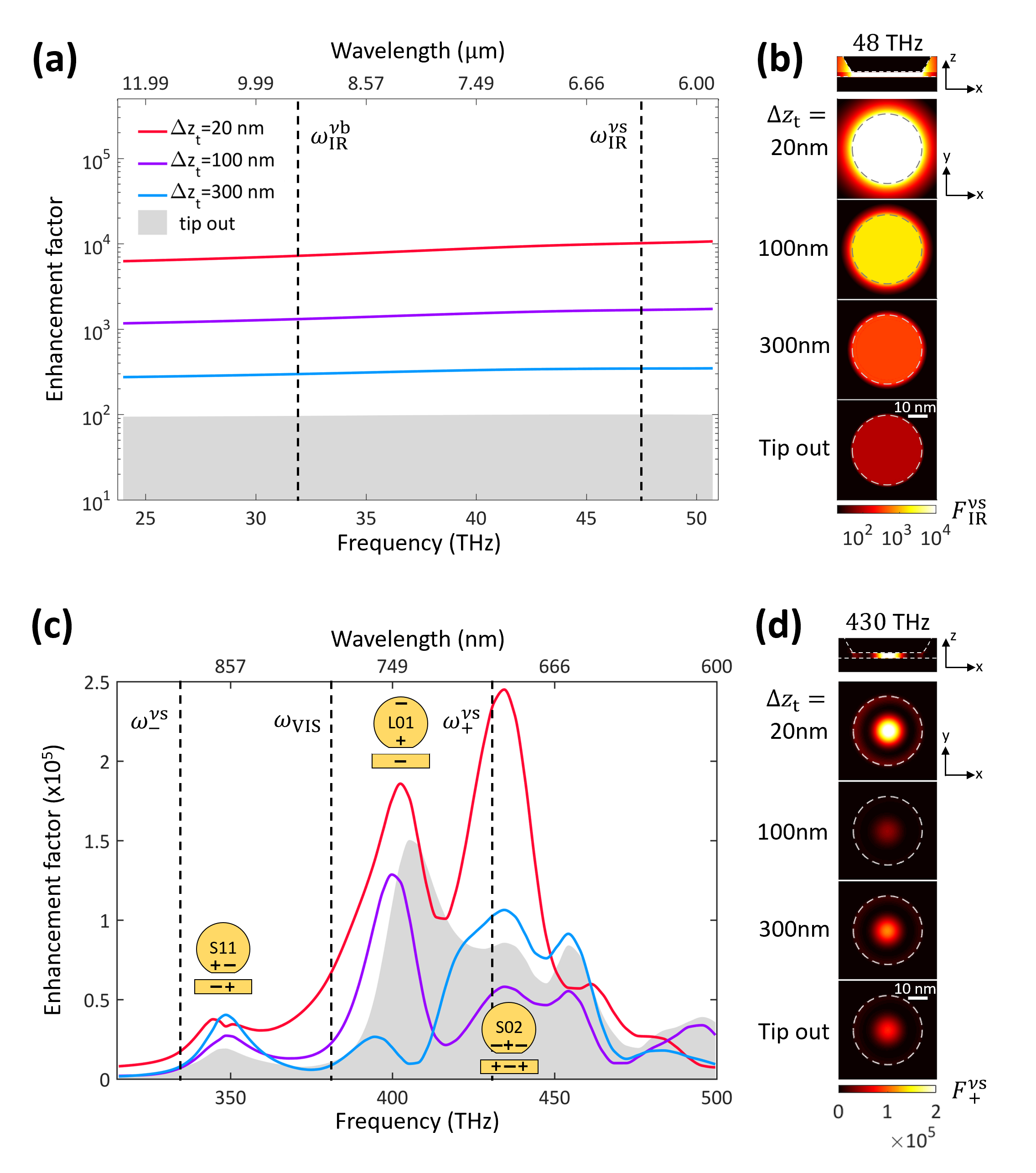}
  \caption{\textbf{Multispectral tip-based nanolens}. 
  Spectra of the simulated maximum intensity enhancement factor inside the NPoM cavity 
  with the Pt tip at position $\mathrm{P1}$ 
  in the infrared (a) and visible (c) frequency ranges for different tip-NPoM distances. 
  Spectra of the NPoM cavity in absence of a tip are depicted by a grey area. 
  Charge distributions of the NPoM cavity modes $S_{\mathrm{11}}$, $L_{\mathrm{01}}$ and $S_{\mathrm{02}}$ 
  are sketched near their resonance frequencies. 
  Panels (b,d) depict the spatial distribution of the intensity enhancement factor at different frequencies: 
  $\omega_{\mathrm{IR}}^{\nu_{\mathrm{s}}}/(2\pi)=48 $~THz in (b) and $\omega_{\mathrm{+}}^{\nu_{\mathrm{s}}}/(2\pi)=430 $~THz in (d). 
  First rows show a vertical cut of the distribution of the intensity enhancement factor inside the NPoM gap 
  for a tip at a distance of $\Delta z_{\mathrm{t}}=20$~nm from the NPoM cavity. 
  Following rows show its distribution on the mid-gap horizontal plane of the NPoM gap 
  for tip-NPoM distances of $\Delta z_{\mathrm{t}}=\{20,\: 100$,\: $300\}$~nm and for the NPoM cavity without tip case. 
  }
  \label{fig:te-gaps-num}
\end{figure}

To verify and better understand the influence of the tip on both the linear and nonlinear optical signals, 
we calculate (see Fig.~\ref{fig:hs-sfg}e) the intensity enhancement distributions inside the NPoM gap, $F(\vec{\mathrm{r}})$, 
for various tip-NPoM distances $\Delta z_{\mathrm{t}}$ at IR (Fig.~\ref{fig:te-gaps-num}b) and VIS (Fig.~\ref{fig:te-gaps-num}d) frequencies. 
Plotting the maxima of intensity enhancement factors, $F(\vec{\mathrm{r}}_{\mathrm{hs}})$, 
as a function of frequency (Fig.~\ref{fig:te-gaps-num}a), 
we observe an increase of about two orders of magnitude across the whole mid-IR spectral range 
when the tip approaches the NPoM cavity, which we attribute to the lightning rod effect. 
The highly efficient IR near-field illumination of the NPoM cavity 
thus explains the remarkable boost of the SFG and DFG signals 
at the frequencies of the molecular vibrational modes $\nu_{\mathrm{b}}$ and $\nu_{\mathrm{s}}$. 
In the visible spectral range, the spectral behavior of $F(\vec{\mathrm{r}}_{\mathrm{hs}})$ is much richer, 
owing to the presence of the different plasmonic modes supported by the NPoM cavity 
(gray area in Fig.~\ref{fig:te-gaps-num}c). 
In addition to the longitudinal dipolar antenna mode (see inset, labelled $L_{\mathrm{01}}$), 
the faceted nanoparticle supports transversal metal-insulator-metal gap modes of different orders and azimuthal symmetry (see inset, labeled $S_{\mathrm{mn}}$) \cite{tserkezis_hybridization_2015,zhang_surface_2019}. 
When the tip is approached towards the NPoM cavity, we observe a frequency-dependent non-monotonic behavior of $F(\vec{\mathrm{r}}_{\mathrm{hs}})$ 
for intermediate distances $40$~nm~$ \lesssim \Delta z_{\mathrm{t}} \lesssim 300$~nm, 
which we attribute to the interference between the tip's near fields and the far-field illumination 
at the position of the NPoM cavity (see Supplementary Note~10). 
For that reason, vibrational signals may not increase monotonically with decreasing $\Delta z_{\mathrm{t}}$, 
as for example observed in Fig.~\ref{fig:control-conversion}c for the non-driven Stokes signal $I^{\nu_{\mathrm{b}}}_{-}$. 
More importantly, for distances $\Delta z_{\mathrm{t}} \lesssim 40$~nm, we find a strong increase of $F(\vec{\mathrm{r}}_{\mathrm{hs}})$ for most frequencies. 
This explains why all Stokes and anti-Stokes signals observed in Fig.~\ref{fig:control-conversion} are larger as the tip approaches, 
independent of whether they are driven by IR fields or not. 

\newpage

\begin{figure}[ht!]
\centering
\vspace*{-15pt}
\includegraphics[scale=0.65]{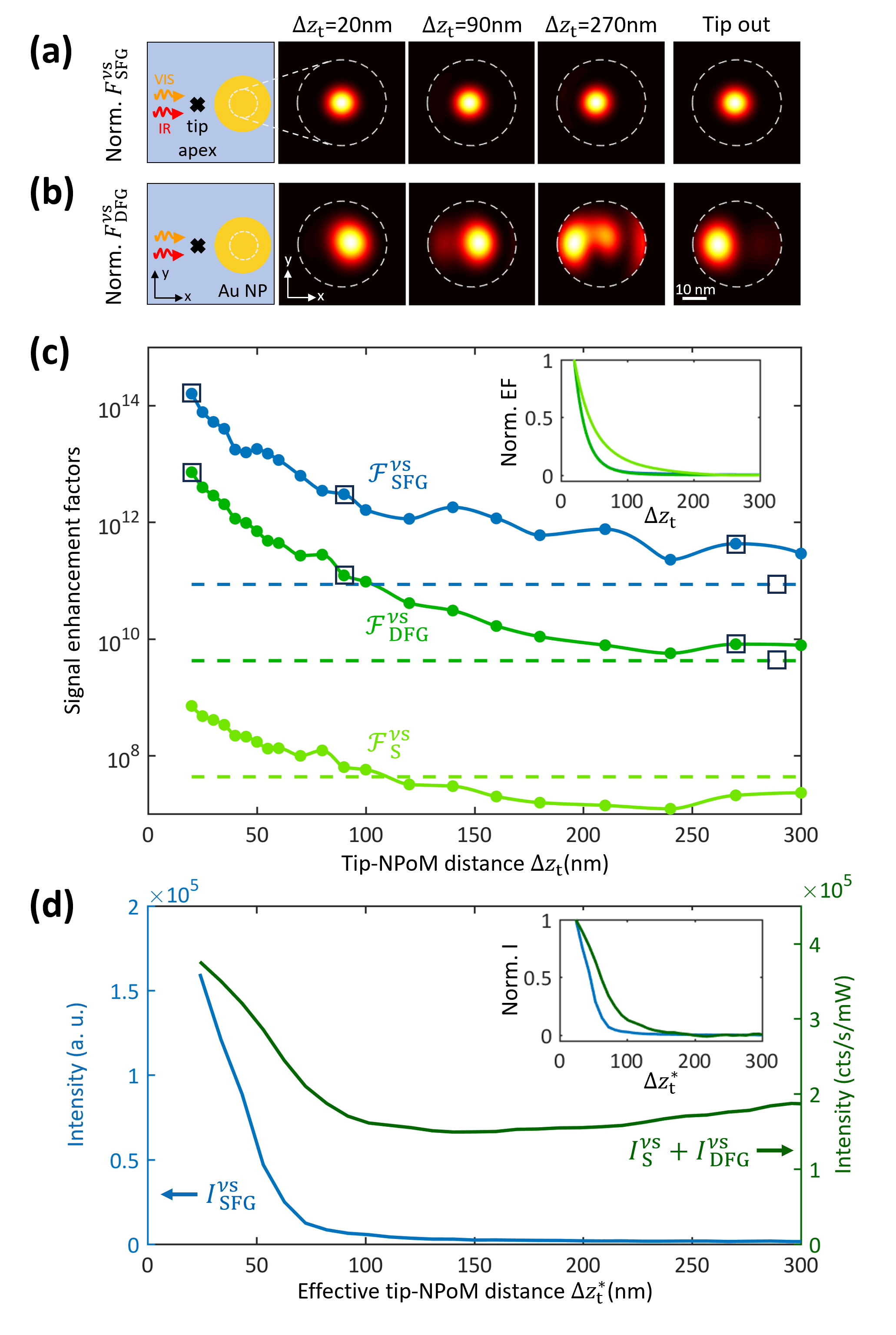}
\vspace*{-15pt}
  \caption{ \textbf{Scaling of tip-enhanced optical signals with tip-NPOM distance}.    
  Numerical calculations of normalized SFG (a) and DFG (b) hot-spots 
  on the mid-gap horizontal plane of the NPoM gap
  for different tip-NPoM distances $\Delta z_{\mathrm{t}}$. 
  Bottom facet of the gold nanoparticle is depicted by a white dashed line. 
  First column displays a top-view schematics of the tip-NPoM configuration and the illumination considered.
  (c) Simulated tip-height dependence of 
  SFG ($\mathcal{F}_{\mathrm{SFG}}^{\nu_{\mathrm{s}}}(\vec{\mathrm{r}}_{\mathrm{hs}})$, blue curve), 
  DFG ($\mathcal{F}_{\mathrm{DFG}}^{\nu_{\mathrm{s}}}(\vec{\mathrm{r}}_{\mathrm{hs}})$, green) 
  and Stokes ($\mathcal{F}_{\mathrm{S}}^{\nu_{\mathrm{s}}}(\vec{\mathrm{r}}_{\mathrm{hs}})$, light green) 
  enhancement factors at the frequency of the vibrational mode $\nu_{\mathrm{s}}$. 
  Inset illustrates the three normalized enhancement factors 
  after smoothing the numerical results with a Whittaker algorithm.
  (d) Sidebands signals (area under the peaks) extracted from the spectra shown in Fig.~\ref{fig:control-conversion}(b) 
  as a function of effective tip-NPoM distance $\Delta z_{\mathrm{t}}^*$. 
  Signal intensity measured on the driven $\nu_{\mathrm{s}}$ upper sideband 
  is shown in blue ($I_{\mathrm{SFG}}^{\nu_{\mathrm{s}}}$). 
  The intensity measured on the $\nu_{\mathrm{s}}$ lower sideband is shown in dark green 
  ($I_{\mathrm{S}}^{\nu_{\mathrm{s}}}+I_{\mathrm{DFG}}^{\nu_{\mathrm{s}}}$). 
  Inset illustrates the normalized behavior of the measured signals after linear background suppression.}
  \label{fig:tbac}
\end{figure}

From the intensity enhancement factor distributions inside the NPoM gap, 
we calculate the spatial distributions of $\mathcal{F}_{\mathrm{SFG}}^{\nu_{\mathrm{s}}}(\Vec{r})$, 
$\mathcal{F}_{\mathrm{DFG}}^{\nu_{\mathrm{s}}}(\Vec{r})$ (Fig.~\ref{fig:tbac}a,b) and $\mathcal{F}_{\mathrm{S}}^{\nu_{\mathrm{s}}}(\Vec{r})$
to determine $\mathcal{F}_{\mathrm{SFG}}^{\nu_{\mathrm{s}}}(\vec{\mathrm{r}}_{\mathrm{hs}})$, 
$\mathcal{F}_{\mathrm{DFG}}^{\nu_{\mathrm{s}}}(\vec{\mathrm{r}}_{\mathrm{hs}})$ 
and $\mathcal{F}_{\mathrm{S}}^{\nu_{\mathrm{s}}}(\vec{\mathrm{r}}_{\mathrm{hs}})$  
as a function of tip-NPoM distance $\Delta z_{\mathrm{t}}$. 
$\mathcal{F}_{\mathrm{SFG}}^{\nu_{\mathrm{s}}}(\vec{\mathrm{r}}_{\mathrm{hs}})$ increases by about two orders of magnitude 
for decreasing $\Delta z_{\mathrm{t}}$ (blue curve in Fig.~\ref{fig:tbac}c), 
reaching a remarkable value of about $10^{14}$ for $\Delta z_{\mathrm{t}}=20$~nm,  
whereas the spatial distribution $\mathcal{F}_{\mathrm{SFG}}^{\nu_{\mathrm{s}}}(\vec{\mathrm{r}}_{\mathrm{hs}})$ remains unchanged 
as compared to the NPoM without tip (Fig.~\ref{fig:tbac}a). 
A similarly strong increase is observed for $\mathcal{F}_{\mathrm{DFG}}^{\nu_{\mathrm{s}}}(\vec{\mathrm{r}}_{\mathrm{hs}})$ 
(green curve in Fig.~\ref{fig:tbac}c), reaching a value about $10^{13}$ for small $\Delta z_{\mathrm{t}}$. 
However, the spatial distribution of $\mathcal{F}_{\mathrm{SFG}}^{\nu_{\mathrm{s}}}(\vec{\mathrm{r}}_{\mathrm{hs}})$ changes 
with decreasing $\Delta z_{\mathrm{t}}$ (Fig.~\ref{fig:tbac}b), showing that the tip can not only be used for enhancing signals 
, but also for controlling the position of hot spots within the NPoM gap. 
In the future, and in case of heterogeneously filled NPoM cavities, such hot-spot control could be used for in-operando selection of specific areas, 
or even single molecules, to which the cavity couples to.  

We highlight that the decrease of SFG and DFG signal enhancement factors 
with increasing tip-NPoM distance $\Delta z_{\mathrm{t}}$ is equal, 
but steeper than the Stokes signal enhancement factor 
(see normalized curves in inset of Fig.~\ref{fig:tbac}c). 
This key advantage of nonlinear signals (e.g. for better spatial localization of an optical response 
\cite{kawata_feasibility_1999,zayats_apertureless_2000,sanchez_near-field_1999}) 
is demonstrated experimentally in Fig.~\ref{fig:tbac}d, where we plot the measured intensities 
$I_{+}^{\nu_{\mathrm{s}}}= I_{\mathrm{SFG}}^{\nu_{\mathrm{s}}} + I_{\mathrm{aS}}^{\nu_{\mathrm{s}}} \sim I_{\mathrm{SFG}}^{\nu_{\mathrm{s}}}$ and 
$I_{-}^{\nu_{\mathrm{s}}}= I_{\mathrm{DFG}}^{\nu_{\mathrm{s}}} + I_{\mathrm{S}}^{\nu_{\mathrm{s}}} + I_{\mathrm{-,\: bg}}^{\nu_{\mathrm{s}}}$
as a function of the effective tip-NPoM distance $\Delta z_{\mathrm{t}}^*$ (see Methods). 
After subtraction of the background signal from the NPoM cavity, $I_{\mathrm{-,\: bg}}^{\nu_{\mathrm{s}}}$, 
and normalization of $I_{+}^{\nu_{\mathrm{s}}}$ and $I_{-}^{\nu_{\mathrm{s}}}$ (inset of Fig.~\ref{fig:tbac}d), 
we clearly see that $I_{+}^{\nu_{\mathrm{s}}}$ decays much stronger with increasing $\Delta z_{\mathrm{t}}^*$ 
than $I_{-}^{\nu_{\mathrm{s}}}$, demonstrating a different scaling of both signals: 
$I_{+}^{\nu_{\mathrm{s}}}$ is a purely nonlinear SFG signal, 
whereas $I_{-}^{\nu_{\mathrm{s}}}$ comprises both a nonlinear DFG and linear Stokes signals. 
We also highlight the increase of $I_{\mathrm{SFG}}^{\nu_{\mathrm{s}}}$ by about two orders of magnitude 
when the tip is approached to the NPoM cavity (blue curve in Fig.~\ref{fig:tbac}d), 
being in good agreement with the calculated SFG signal enhancement factor 
$F_{\mathrm{SFG}}^{\nu_{\mathrm{s}}}(\vec{\mathrm{r}}_{\mathrm{hs}})$ (blue curve in Fig.~\ref{fig:tbac}c).

Fig.~\ref{fig:tbac} demonstrates that tip-enhanced nonlinear SFG signals offer distinct advantages compared to tip-enhanced linear Raman signals. 
We find a significantly stronger signal enhancement factor, paired with a more pronounced decay as the tip-sample distance increases, 
which requires a closer proximity of the tip to the object (in our study the NPoM cavity) for signal acquisition. 
Further, we do not observe SFG background signals generated by the far-field illumination. 
We thus envision tip-enhanced continuous wave SFG spectroscopy for background-free nanoimaging of molecules or 2D materials on bare substrates. 
This could be achieved with scanning probe tips providing sufficiently large field enhancement 
at their apex at both visible and infrared frequencies, eliminating the need for NPoM cavities.

\newpage

\section{Discussion}
In summary, we have introduced in-operando control of SFG from molecule-filled NPoM cavities 
via the tip of a scattering-type scanning near-field optical microscope (s-SNOM), 
adding a versatile technique to the toolbox of nonlinear nanooptics 
\cite{yin_edge_2014,kravtsov_plasmonic_2016,yao_nanoscale_2022,takahashi_broadband_2023}. 
Remarkably, low-power continuous wave illumination at both visible and mid-infrared frequencies 
is sufficient to access second-order nonlinearities of individual molecule-filled nanocavities
, which in future could be applied for studying molecular self-assembled monolayers domains \cite{gray_2d_2021} 
or even for characterizing the anharmonicities of molecular vibrations  
and their intramolecular vibrational relaxation mechanisms \cite{chen_sommerfeld_2019,chen_cavity-enabled_2022}. 
Our approach thus circumvents various typical challenges of nonlinear nanooptics, 
such as establishing precise temporal overlap of short high-power laser pulses. 
Further, our prototypical realization of a cascaded nanolens \cite{li_self-similar_2003} 
allows for deep sub-wavelength field concentration across the whole visible to terahertz spectral range, 
greatly relaxing the requirements on samples that can be studied via nonlinear spectroscopy. 
In combination with extreme near-field concentration provided by atomic scale protrusions inside the NPoM gap \cite{baumberg_picocavities_2022}, 
SFG involving dipole-forbidden contributions may become a promising topic of research. 
In the future, optimized scanning probe tips exhibiting simultaneously strong resonances at both IR and visible frequencies 
may pave the way to v-SFG spectroscopy without the need of NPoM cavities and thus for v-SFG nanoimaging.

\newpage

\section*{Materials and methods} 
\paragraph{Setup description:}
A schematic of the setup and a detailed description of its elements is provided in Supplementary Note~1. 
We use a commercial s-SNOM (neascope, attocube AG) is based on an AFM,  whose metallic tip is used to 
enhance the incident VIS and IR fields at its apex (nano-FTIR tips with nominal tip apex diameter of 100~nm). 
In proximity to an NPoM cavity, the tip's near field provides a strongly concentrated illumination in addition to the incident far-field illumination.  
In our experiment, the sample and tip are illuminated from the side with radiations from CW monochromatic IR and VIS laser sources. 
The visible beam is first spectrally filtered with a laser line bandpath filter in order to allow for Raman and photoluminescence (PL) spectroscopy measurements. 
Both collimated beams pass then through beam expanders (reaching 1~cm beam diameter) and attenuators before being combined on a dichroic plate 
and sent to the high-NA off-axis parabolic mirror (PM) or our s-SNOM. 
The PM tightly focuses the laser beams onto the oscillating AFM tip (frequency $\Omega\sim250$~kHz, 
tapping amplitude $\mathrm{TA} \sim 50$~nm) and collects the backscattered light. 
To precisely focus both beams onto the tip apex, we record the IR and VIS elastically scattered fields using a MCT detector and a photomultiplier tube (PMT), respectively, in the same fashion as in self-homodyne s-SNOM operation. That is, we optimize the detector signal that is recorded without interferometer and demodulated at the third or fourth harmonic of the tip oscillation frequency, $3\Omega$ or $4\Omega$, respectively. 
After both VIS and IR s-SNOM signals are maximized, the mirror in front of the PMT is flipped such that the back-scattered visible light is guided to a grating spectrometer for Raman or PL measurements. 
The VIS excitation and radiation from the AFM deflection laser are removed by placing corresponding filters (NF/LP+SP) in front of the spectrometer entrance.  

\paragraph{Tip-based approach curves (TBAC):}
In contrast to standard s-SNOM, the oscillating tip rather than the sample 
is retracted/approached in our experiments to ensure that changes of the optical signals 
are not caused by movement of the NPoM cavity within the focus of the far-field illumination. 
To that end, we apply a linear voltage ramp to the piezo (in the AFM head) onto which the AFM cantilever is mounted. 
This linear voltage ramp is applied as an offset voltage additional 
to the sinusoidal voltage for exciting the cantilever (tip) oscillation. 
For converting the piezo voltage $U_t$ into tip-sample distance $\Delta z_{\mathrm{tbac}}$, 
we record a TBAC on the Au mirror next to the AuNP 
and a standard approach curve using the calibrated sample scanner. 
Comparison of both approach curves (recorded with the same parameters and alignment) 
allows for converting $U_t$ to $\Delta z_{\mathrm{tbac}}$. 
The corresponding microscope script is available in the Zenodo repository.

\paragraph{Effective tip-NPoM distance $\Delta z^{*}_{\mathrm{t}}$:} 
In order to compare the experimental results obtained with a tip oscillating at tapping amplitude $\mathrm{TA}=50$~nm 
and the numerical calculations performed with a static tip, 
we introduce the offset distance $z^{*}_{\mathrm{t,0}}=20$~nm. 
The offset is chosen such that numerical calculations for a tip-NPoM distance of $z^{*}_{\mathrm{t,0}}$ reproduce faithfully the spatial variations of optical signals 
$I_{\mathrm{SFG}}$, $I_{\mathrm{DFG}}$, $I_{\mathrm{aS}}$ and $I_{\mathrm{S}}$ observed experimentally with an oscillating tip in contact mode. The effective tip-NPoM distance can consequently be defined as: 
$\Delta z^{*}_{\mathrm{t}} = \Delta z_{\mathrm{tbac}} + z^{*}_{\mathrm{t,0}}$, 
with $\Delta z_{\mathrm{tbac}}$ the variation of distance recorded during a TBAC (see section above). 

\paragraph{Sample preparation:}
A detailed description of the different sample preparation steps can be found elsewhere \cite{chen_intrinsic_2021}. 
In short, template-stripped gold films are incubated for 3 hours in a biphenyl-4-thiol (BPT) solution (ethanol buffer, Sigma-Aldrich), 
yielding a self-assembled monolayer (SAM) of BPT. 
Subsequently, faceted gold nanoparticles 
(BBI solutions, 80~nm nominal diameter) are drop-casted on the BPT functionalized gold films. 
After incubation of 10 to 30 min (depending on the concentration of NP solution), the samples are gently rinsed with DI water and cleaned with nitrogen gas. 
Far-field optical characterization of the obtained NPoM samples
show that the particles strongly enhance the Raman signal of the molecules, 
demonstrating that they act as nanocavities (Supplementary Note~2).

\paragraph{Numerical simulations:}
The electromagnetic simulations are performed with a commercial FEM package (COMSOL). The simulation universe contains a Pt tip, 
modelled as a $1~\mu$m height half ellipsoid with 100~nm apex diameter, 
a gold nanoparticle with a facet size of $w=35$~nm and a height of $h=60$~nm, which is separated from a 150~nm thick gold film by a $d=1.15$~nm thick dielectric layer with refractive index $n=1.4$ (see Supplementary Note~7 of SI for a justification of the choice of the tip height and tip apex geometry). 
The combination of optical gap size ($n\cdot d$) and particle's facet size ($w$) are chosen 
to fit well the AFM and photoluminescence signals of representative NPoM cavities 
under our microscope scanning tip (cf. Supplementary Note~9 in SI).
The simulation universe is surrounded by a 400~nm thick PML layer. 
Special care is taken to mesh the contact regions of our cascaded nanostructure. 
The dielectric function of gold is obtained by linearly interpolating 
the data provided in Ref.~\cite{johnson_optical_1972} for the VIS spectral range 
and in Ref.~\cite{babar_optical_2015} for the IR range. 
For simplicity, we consider intensity enhancement factors given by $F=(|E_s|/|E_0|)^2$ 
instead of $F=(1+E_s/E_0)^2$ \cite{novotny_principles_2012} as the scattered fields $E_s$ in the NPoM gap 
are strongly dominated by the out-of-plane component and much greater than the incident field $E_s \gg E_0$. 
$|E_s|$ is numerically calculated for a plane-wave excitation $E_0$ incident at an angle of 35 degrees from the surface.

\section*{Acknowledgements}
The authors thank Monika Goikoetxea for her assistance with the samples, Carlos Crespo and Andrei Bylinkin 
for their assistance with the optical setup, Mario Zapata for his help with the simulations and Andrey Chuvilin for taking the SEM image of the tip. 
The work received support from Grant CEX2020-001038-M funded by MICIU/AEI /10.13039/501100011033 
and Grant PID2021-123949GB-I00 (NANOSPEC) funded by MICIU/AEI/10.13039/501100011033 and by ERDF/EU. 
P. R. acknowledges financial support from the Swiss National Science Foundation (Grant No. 206926)
and from the European Union's Horizon 2020 research and innovation programme under the Marie Skłodowska-Curie grant agreement No 101065661. 
I. N. acknowledges financial support from the Ministerium für Innovation, Wissenschaft und Forschung des Landes Nordrhein-Westfalen as well as from the Deutsche Forschungsgemeinschaft (project number 467576442). 
JA and IP acknowledge funding from project IT1526-22 from the Department of Education of the Basque Government, from Elkartek project "u4smart" from the Dept. of Industry of the Basque Government, and from grant PID2022-139579NB-I00 funded by MCIN/AEI/10.13039/501100011033 and by “ERDF A way of making Europe”. 


%

\widetext
\clearpage

\setcounter{equation}{0}
\setcounter{figure}{0}
\setcounter{table}{0}
\makeatletter
\renewcommand{\thesubsection}{\textbf{Supplementary Note \arabic{subsection}}}
\renewcommand{\figurename}{\textbf{Supplementary Figure}}
\renewcommand{\tablename}{\textbf{Supplementary Table}}
\renewcommand{\thefigure}{S\arabic{figure}}
\renewcommand{\theequation}{S\arabic{equation}}
\renewcommand{\thetable}{S\arabic{table}}
\renewcommand*{\citenumfont}[1]{S#1}
\renewcommand*{\bibnumfmt}[1]{[S#1]}


\subsection{Optical setup}
\begin{figure}[ht!]
\centering
\vspace*{-18pt}
\includegraphics[width=0.65\textwidth]{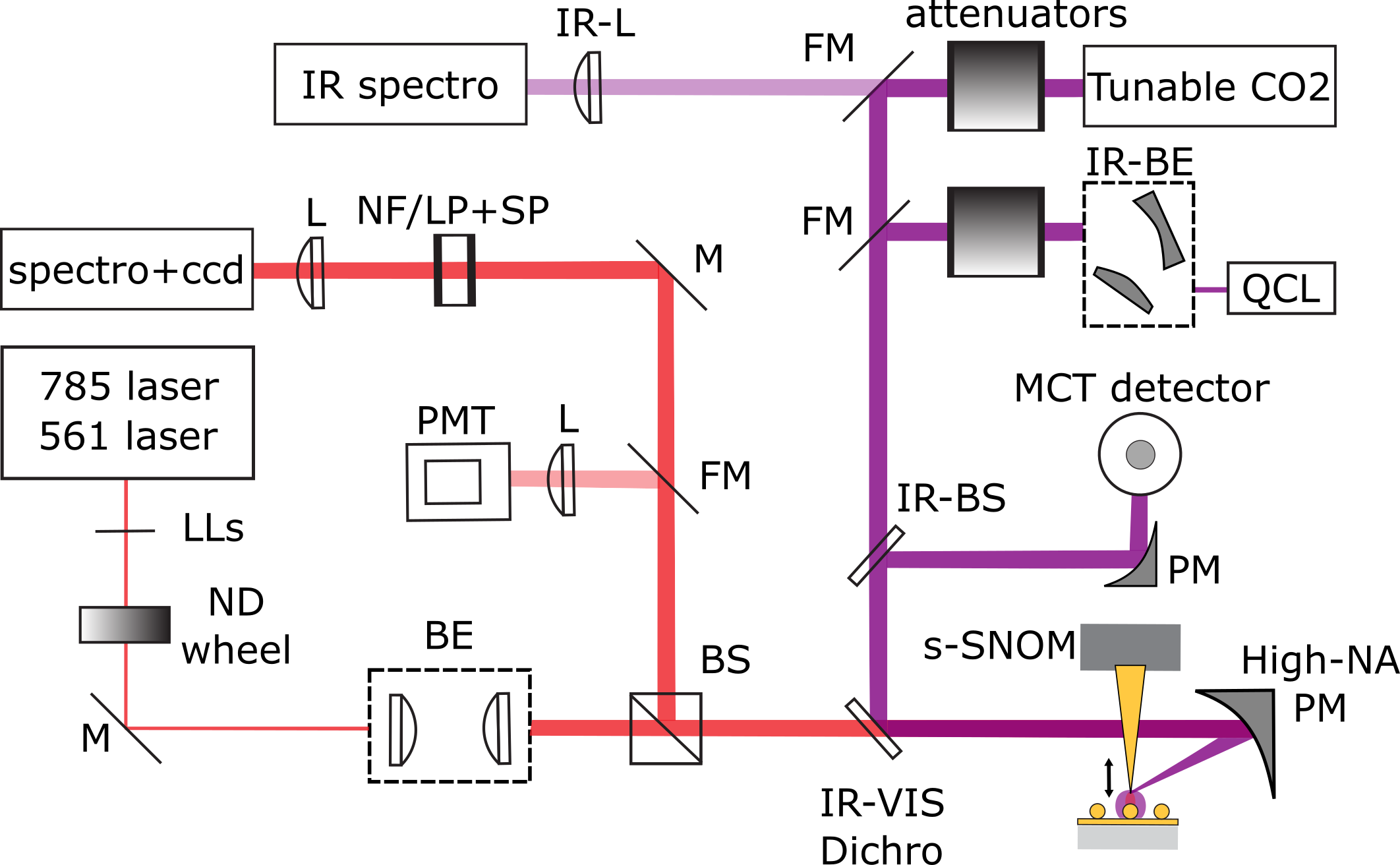}
\vspace*{-10pt}
  \caption{
  \textbf{Schematic of the optical setup}. 
  \textit{VIS tip illumination path.} 
  \textbf{785 laser}: 
  Raman excitation via 785~nm CW single mode wavelength-stabilized laser diode
  (SureLock785 from Coherent), 
  \textbf{561 laser}: 
  PL excitation via 561~nm CW single wavelength laser diode
  (Cobolt 08-DPL from H\"ubner Photonics), 
  \textbf{LLs}: 
  laser line bandpass filters for the 785~nm 
  (LL01-785-12.5)
  and 561~nm lasers
  (LL02-561-12.5 from Semrock), 
  \textbf{ND wheel}: 
  neutral density filter wheel, 
  \textbf{BE}: 
  Keplerian type beam expander with additional spatial filtering, 
  \textbf{BS}: 
  90:10 (R:T[\%]) non-polarizing broadband cube beamsplitter, 
  \textbf{IR-VIS Dichro}: 
  ITO front and AR back coated glass plate used as visible infrared dichroic mirror 
  (from Newport).   
\textit{IR tip illumination path.} 
  \textbf{Tunable CO2}: 
  9.2-10.8~$\mu$m adjustable wavelength CO$_2$ CW monochromatic gas laser 
  (Merit-G from Access Laser Company), 
  \textbf{QCL}: 
  5.7-6.8~$\mu$m tunable external cavity IR CW monochromatic mode-hop-free QCL laser  
  (Tunable Laser System from Daylight Solutions), 
  \textbf{IR-BE}: 
  reflective beam expander 
  (BE04 from Thorlabs), 
  \textbf{attenuators}: 
  metal grid based variable attenuator for the CO${_2}$ laser path 
  (IR attenuator from Lasnix) 
  and double broadband IR linear polarizers variable attenuator for the QCL path, 
  \textbf{IR-BS}: 
  50:50 (R:T[\%]) ZnSe broadband beamsplitter. 
\textit{Scanning probe setup.} 
  \textbf{s-SNOM}:
  customized scattering-type scanning near-field microscope 
  (NeaSNOM from attocube), 
  being operated with platinum-iridium tips with 100~nm tip apex diameter
  (nano-FTIR from attocube), 
  \textbf{High-NA PM}: 
  silver protected off-axis parabolic mirror 
  (NA$=0.72$ --- NA$_{\mathrm{eff}}\sim0.55$ for our beam parameters, EFL$ \simeq 6.5$~mm). 
\textit{IR detection path.} 
  \textbf{PM}: 
  silver protected parabolic mirror, 
  \textbf{MCT detector}: 
  Wideband cooled photovoltaic HgCdTe(MCT) detector with amplifier
  (KLD-0.1 from Kolmar Technologies), 
  \textbf{IR spectro}: 
  Analog CO${_2}$ laser spectrometer 
  (CO${_2}$ Laser Spectrum Analyzer from Optical Engineering). 
\textit{VIS detection path.}
  \textbf{PMT}: 
  photomultiplier tube 
  (R9110 from Hamamatsu) 
  and variable gain low noise current amplifier 
  (DLPCA-200 from FEMTO Messtechnik), 
  \textbf{SP+NF/LP}: 
  shortpass filter for AFM deflection laser suppression
  (FF01-945/SP-25) 
  and 785~nm notch filter for Raman measurements
  (NF03-785E-25) 
  or 561~nm longpass filter for PL measurements 
  (BLP02-561R-25 from Semrock), 
  \textbf{spectro+ccd}: 
  Visible grating spectrometer with 300~l/mm grating 
  (Kymera 328i from Andor)
  and CCD camera 
  (DU420A-BEX2-DD from Andor)
\textit{Miscellaneous.}
  \textbf{M}: 
  broadband silver mirror, 
  \textbf{FM}: 
  flip mirror, 
  \textbf{L}: 
  broadband achromatic doublets, 
  \textbf{IR-L}: 
  ZnSe broadband plano-convex lens.
  }
  \vspace*{-24pt}
  \label{fig:sm-setup}
\end{figure}

\newpage

\subsection{Far-field Raman characterization of the NPoM samples}

\begin{figure}[ht!]
\centering
\includegraphics[width=0.95\textwidth]{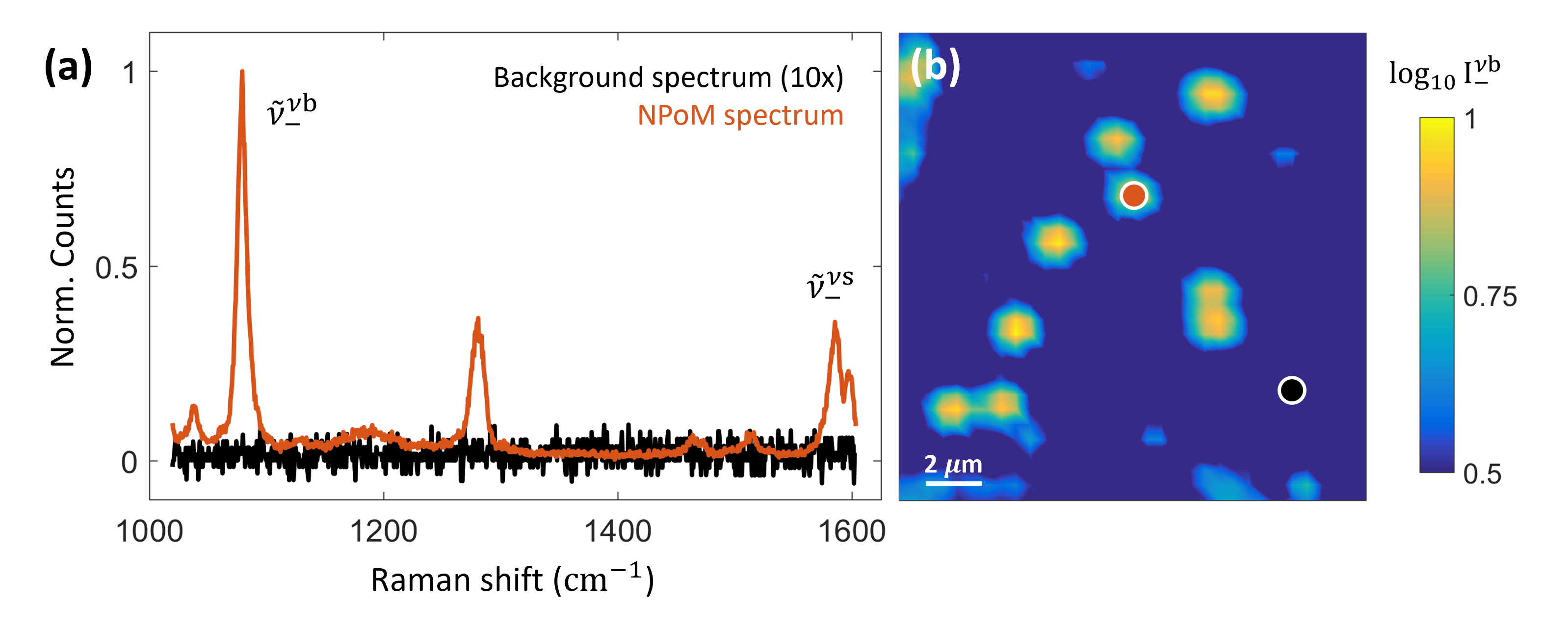}
  \caption{\textbf{Raman signal of BPT molecules inside NPoM nanocavities}. 
  (a) Surface-enhanced Raman spectrum of different vibrational modes of BPT molecules inside a NPoM nanocavity 
  (normalized, red line). 
  The normalized spectrum (10x amplified) in absence of nanocavity is shown in black for comparison. 
  The spatial location corresponding to the two spectra is depicted on the Raman map of panel (b). 
  (b) The Raman map shows the normalized intensities at the $\nu^{\mathrm{b}}_-$ lower vibrational sideband in logarithmic scale. 
  These far-field measurements are performed with a commercial Renishaw microscope 
  (785~nm illumination at a power of 200~$\mu$W,  acquisition time per pixel of 0.5~s, objective NA of 0.85). 
  For the measurements shown in the manuscript, isolated nanocavities are selected to avoid 
  contamination of detected signals from other nearby nanocavities. } 
  \label{fig:sm-raman}
\end{figure}

\newpage

\subsection{Tip-controlled optical signals from a molecule-filled NPoM cavity}

\begin{figure}[ht!]
\centering
\includegraphics[width=0.85\textwidth]{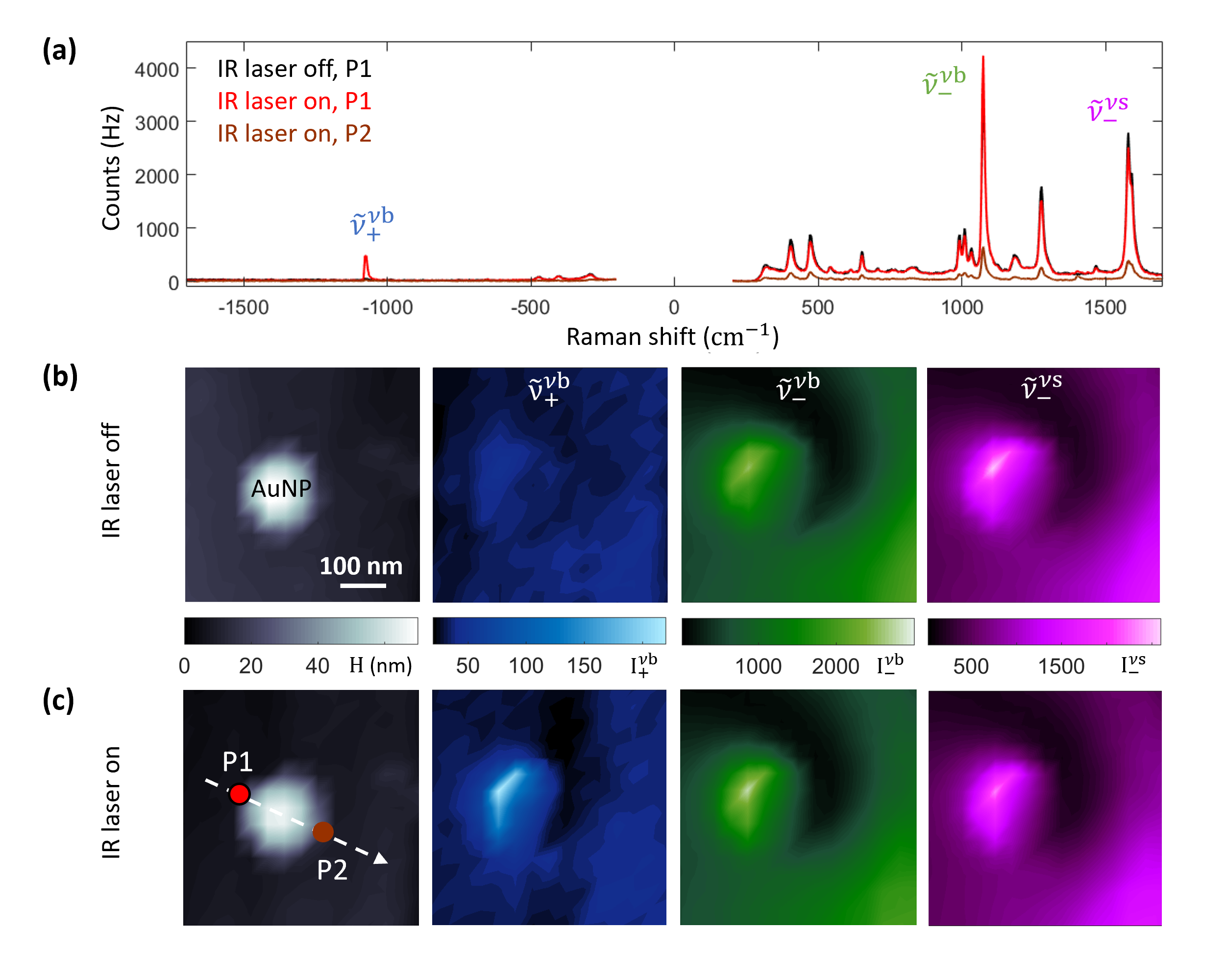}
  \caption{\textbf{Tip-controlled optical signals}.  
  (a) Spectra of a BPT-filled NPoM cavity ($80$~nm nominal diameter Au nanosphere) functionalized with BPT molecules 
  below an oscillating scanning tip ($100$~nm apex diameter).  
  Black spectrum is recorded for tip positioned at position P1 under VIS illumination. 
  Red and brown spectra are recorded under VIS and IR 
  ($2.2$~mW tuned to the vibrational mode $\nu_{\mathrm{b}}$ at 32~THz) 
  illumination at positions P1 and P2, respectively.  
  For a better comparison of the spectra with standard Raman spectra of BPT, 
  we show the spectra in wavenumbers corresponding to the Raman shift. 
  The left part of the red and black spectra are the same data as the ones shown in Fig. 2a.
  (b,c) Simultaneously recorded topography and optical maps 
  without (b) and with (c) IR illumination. 
  Optical maps at the frequencies $\omega_{\mathrm{+}}^{\nu\mathrm{b}}/(2\pi)=414 $~THz 
  (corresponding to a Raman shift of $-1080$~cm$^{-1}$), 
  $\omega_{\mathrm{-}}^{\nu\mathrm{b}}/(2\pi)=350 $~THz ($1080$~cm$^{-1}$) 
  and $\omega_{\mathrm{-}}^{\nu\mathrm{s}}/(2\pi)=335 $~THz ($1575$~cm$^{-1})$
  are shown in blue, green and pink, respectively. 
  Red and brown dots indicate the tip positions P1 and P2 where spectra shown in panel (a) are recorded.
  The illumination direction is illustrated by the white dashed arrow in (c). 
  VIS illumination is at $382$~THz ($785$~nm) at a power of $200$~$\mu$W. 
  Acquisition time per spectrum is $2$~s. Tapping amplitude (TA) is $50$~nm.
  Topography image and optical maps at frequency $\omega_{\mathrm{-}}^{\nu\mathrm{b}}/(2\pi)$ 
  are the same as the ones shown in Fig.2c and Fig.2d of the main text. 
  }
  \label{fig:sm-hs-maps}
\end{figure}

\newpage

\subsection{Tip-controlled SFG signals from other NPoM cavities}

\begin{figure}[ht!]
\centering
\includegraphics[width=1.\textwidth]{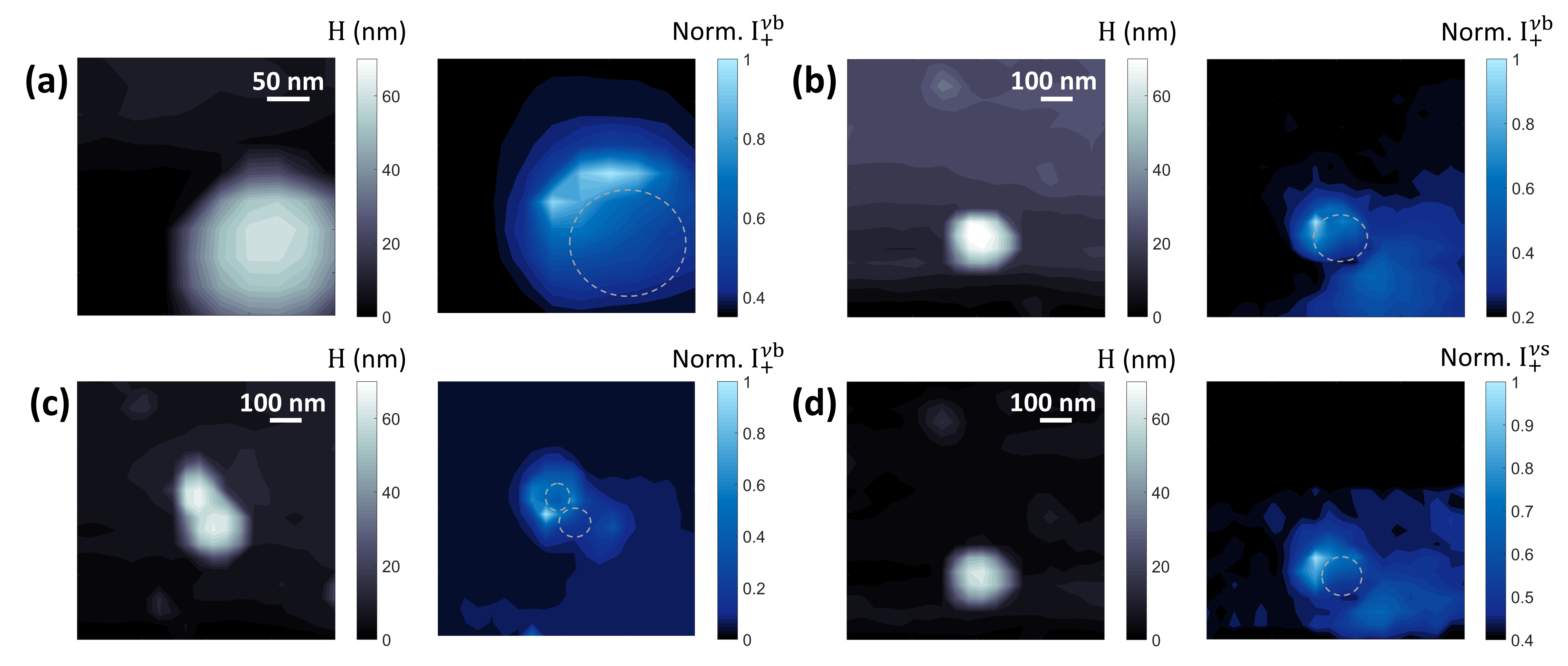}
  \caption{\textbf{Tip-controlled cavity-enhanced SFG}.  
  Optical maps at IR driven upper vibrational frequencies for other representative NPoM cavities of our samples, under similar illumination direction as in Fig.~\ref{fig:sm-hs-maps}. 
  Each panel is composed in the following way. 
  Right: Topography image of the NPoM cavity. 
  Left: Normalized optical map of the signal $I_{\mathrm{+}}$ as a function of tip position. The illumination direction is illustrated by the white dashed arrow. 
  For improved clarity, the contour of the nanoparticle is outlined with a dashed gray line on the optical maps. 
  In panels (a-c), the IR illumination is tuned to the vibrational mode $\nu_{\mathrm{b}}$ at 32~THz 
  and the optical maps are recorded at $\omega_{\mathrm{+}}^{\nu\mathrm{b}}$. 
  In panel (d), the NPoM cavity shown in (b) is illuminated with an IR illumination tuned to the vibrational mode $\nu_{\mathrm{s}}$ at 48~THz and the optical map is recorded at $\omega_{\mathrm{+}}^{\nu\mathrm{s}}$. 
  }
  \label{fig:sm-sfg-maps}
\end{figure}

\newpage

\subsection{Intensity enhancement factors as a function of tip position} 

\begin{figure}[ht!]
\centering
\includegraphics[width=1.\textwidth]{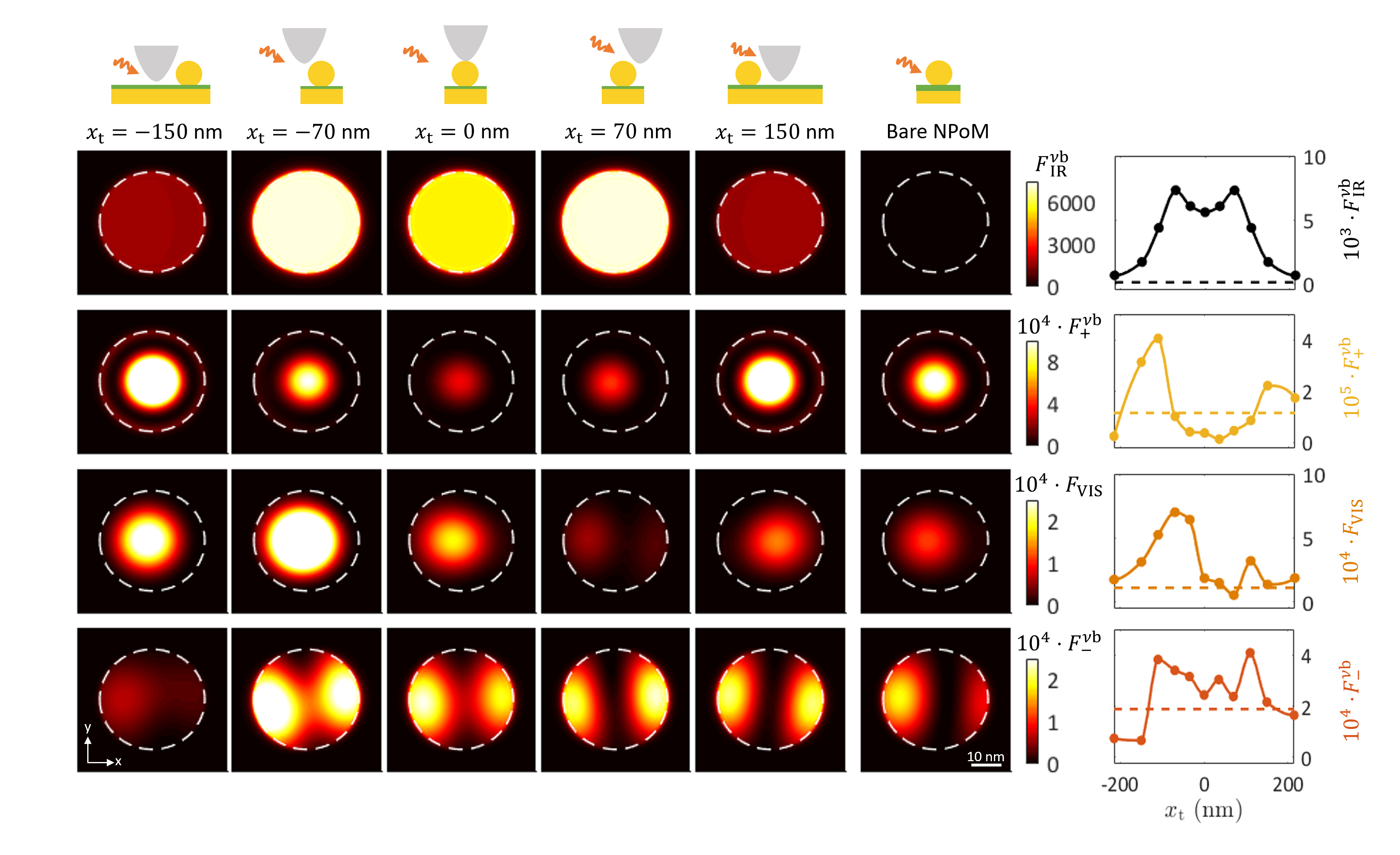}
  \caption{\textbf{Simulated intensity enhancement factor distributions for different tip positions}. 
  Main panel: 
  Intensity enhancement factor spatial distributions on the mid-gap horizontal plane of the NPoM gap at 
  $\omega_{\mathrm{IR}}^{\nu\mathrm{b}}$, $\omega_{\mathrm{+}}^{\nu\mathrm{b}}$, 
  $\omega_{\mathrm{VIS}}$ and $\omega_{\mathrm{-}}^{\nu\mathrm{b}}$
  for tip positions illustrated by the schematics of the top row. 
  $\Delta z_{\mathrm{t}}\simeq20$~nm for all simulations.  
  The bottom facet of the gold nanoparticle is depicted by a white dashed line. 
  Top row: Sketches of the tip position with respect to the nanoparticle. 
  The orange arrow indicates the illumination direction. 
  Wavelength of VIS illumination is 785~nm. 
  Right column: 
  $F(\vec{\mathrm{r}}_{\mathrm{hs}})$ as a function of tip position $x_{\mathrm{t}}$. 
  The dashed horizontal lines depict $F(\vec{\mathrm{r}}_{\mathrm{hs}})$ 
  for the NPoM cavity without tip.   
  }
  \label{fig:sm-comsol-fields}
\end{figure}

\newpage

\subsection{Signal enhancement factors as a function of tip position}

\begin{figure}[ht!]
\centering
\includegraphics[width=1.\textwidth]{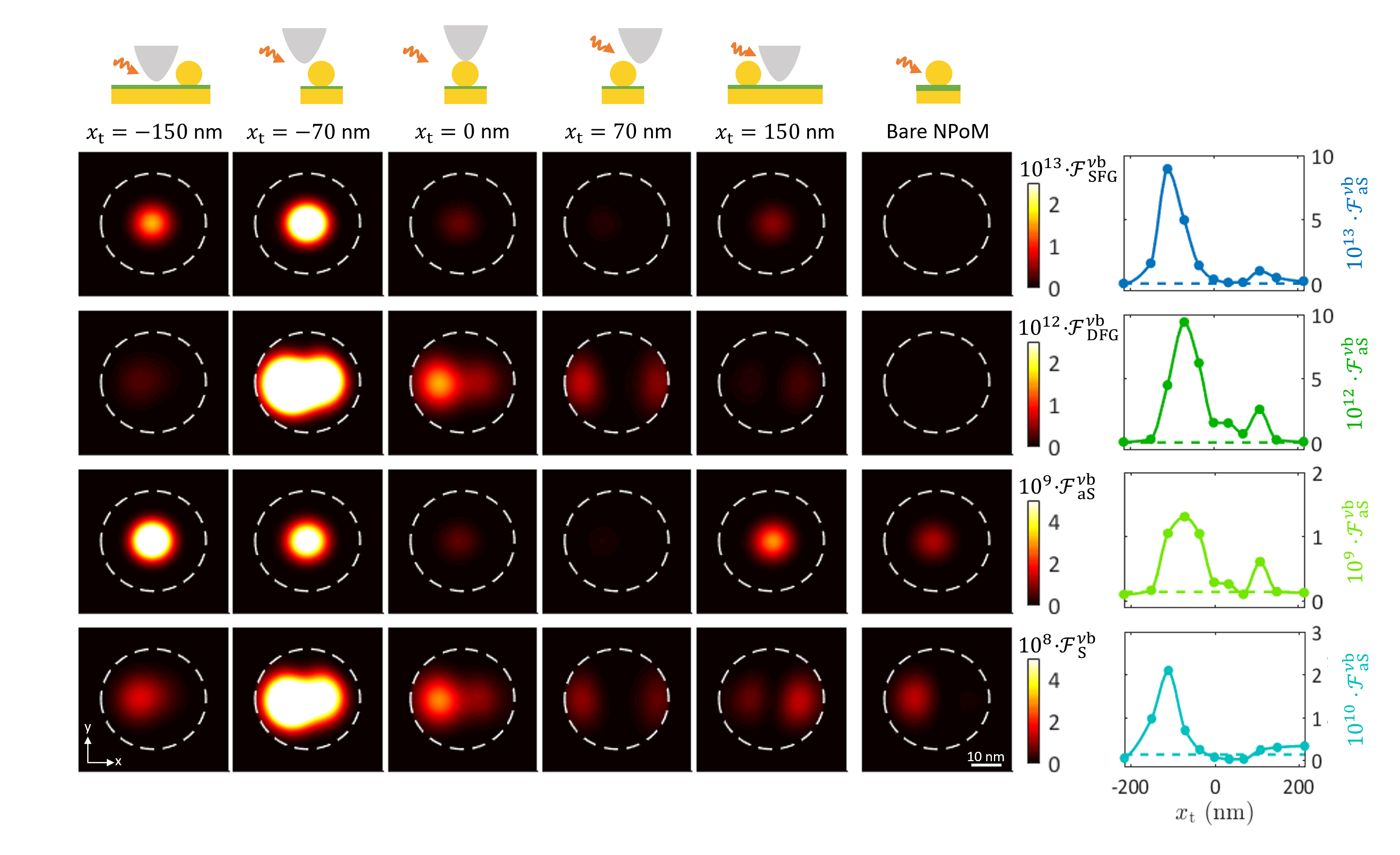}
  \caption{\textbf{Simulated signal enhancement factor distributions for different tip positions}. 
  Main panel: 
  SFG, DFG, aS and S enhancement factor spatial distributions on the mid-gap horizontal plane of the NPoM gap 
  for tip positions illustrated by the schematics of the top row. 
  $\Delta z_{\mathrm{t}}\simeq20$~nm for all simulations. 
  The bottom facet of the gold nanoparticle is depicted by a white dashed line. 
  Top row: Sketches of the tip position with respect to the nanoparticle. 
  The orange arrow indicates the illumination direction. 
  Wavelength of VIS illumination is 785~nm.
  Right column: 
  $\mathcal{F}(\vec{\mathrm{r}}_{\mathrm{hs}})$ as a function of tip position $x_{\mathrm{t}}$. 
  The dashed horizontal lines depict $\mathcal{F}(\vec{\mathrm{r}}_{\mathrm{hs}})$ 
  for the NPoM cavity without tip.
  }
  \label{fig:sm-comsol-signals}
\end{figure}

\newpage

\subsection{Modelling of the metal tip}

\begin{figure}[ht!]
\centering
\vspace{-28pt}
\includegraphics[scale=0.5]{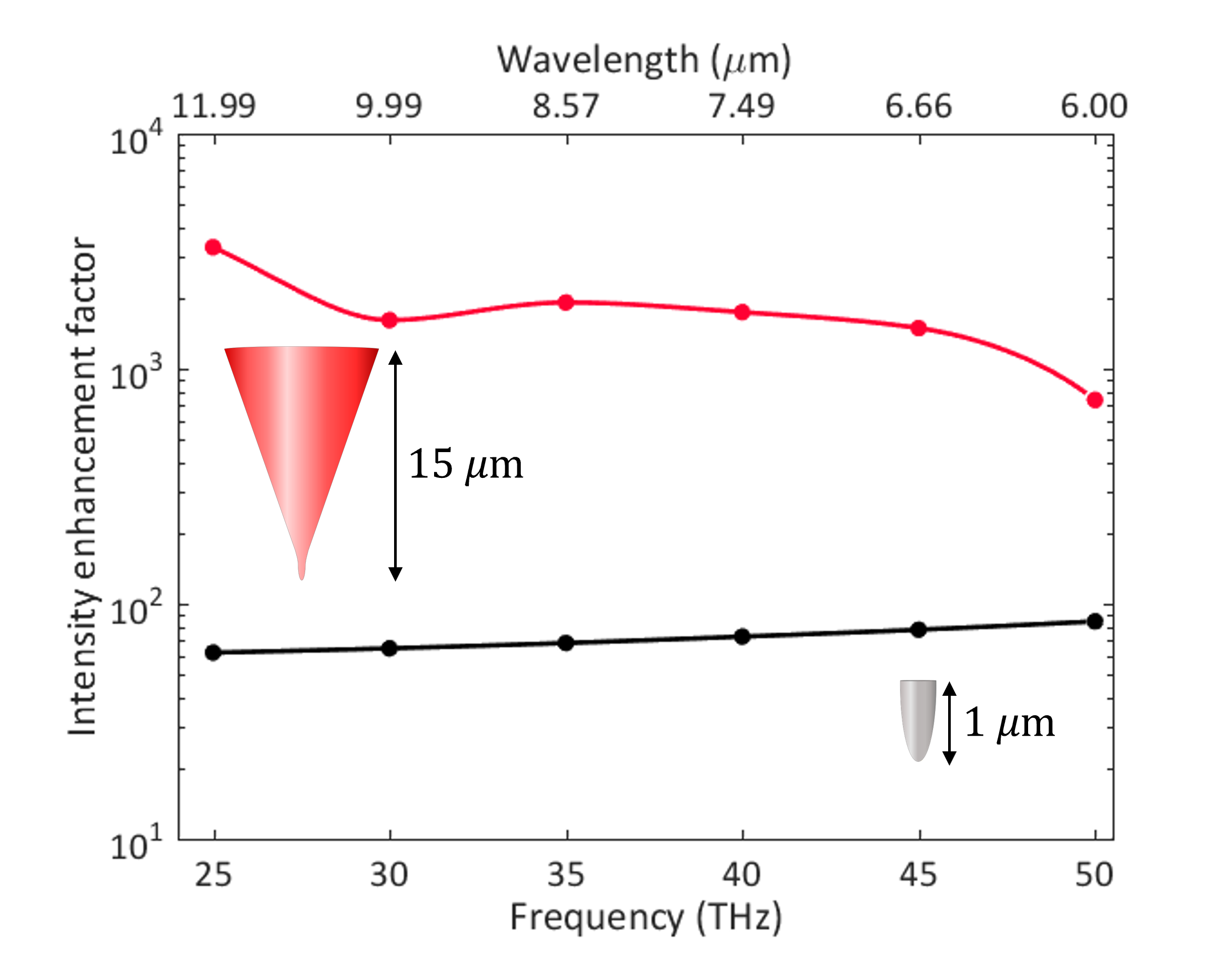}
\vspace{-20pt}
  \caption{\textbf{Simulated intensity enhancement factors for long vs short metal tips}. 
  Numerically calculated spectra of the maximum intensity enhancement factor 
  5~nm below the apex of an isolate tip (i.e. without sample) in the mid-IR frequency range for two tip geometries:  
  the geometry used in the numerical calculations of the manuscript (black line) 
  and a geometry closer to the shape and dimension of the tip used experimentally (red line) \cite{mcardle_near-field_2020}. 
  Sketches illustrate the two different geometries considered. 
  The dimensions of the short tip were chosen so that 
  the heavily multi-scale tip-enhanced NPoM cavity simulations converge. 
  The figure shows that the long tips used in the experiments ($15\ \mu$m long) may have intensity enhancement factors 
  between 8 and 50 times stronger than the short tips used in simulations.  
  The IR enhancement factors discussed in the main text can therefore be considered as a conservative estimate. 
  The slight spectral variation for the longer tip can be attributed to 
  higher order broad resonances of the long conical tip \cite{huth_resonant_2013}.
  }
  \label{fig:sm-tips-ef}
\end{figure}

\begin{figure}[ht!]
\centering
\vspace{-20pt}
\includegraphics[scale=0.55]{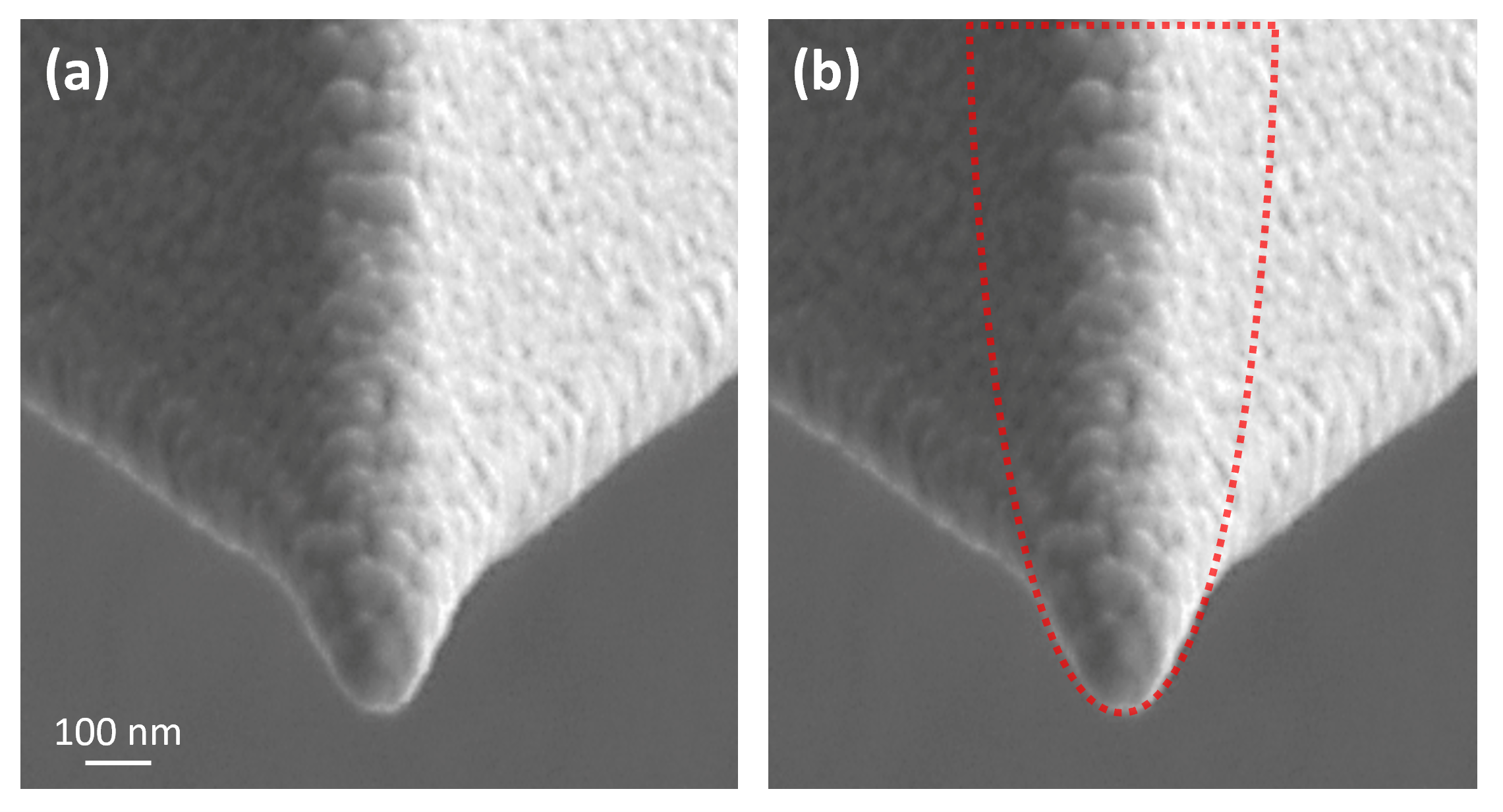}
\vspace{-12pt}
  \caption{\textbf{Comparison of tip geometry in experiment and simulation}. 
  (a) Scanning electron microscopy (SEM) image of one nano-FTIR tip used in the experiment captured at a tilt angle of $52^\circ$. 
  The image is stretched along the y-axis to compensate for the tilt and to accurately depict the proportions of the tip apex.
  (b) The geometry of the tip used in the simulations 
  (a $1\ \mu$m long half-ellipsoid with 100~nm apex diameter)
  is depicted with a dashed red line on top of the SEM image presented in (a).
  }
  \label{fig:sm-tips-sem}
\end{figure}

\newpage

\subsection{Spatial distribution of intensity enhancement around the tip}

\begin{figure}[ht!]
\centering
\includegraphics[scale=0.55]{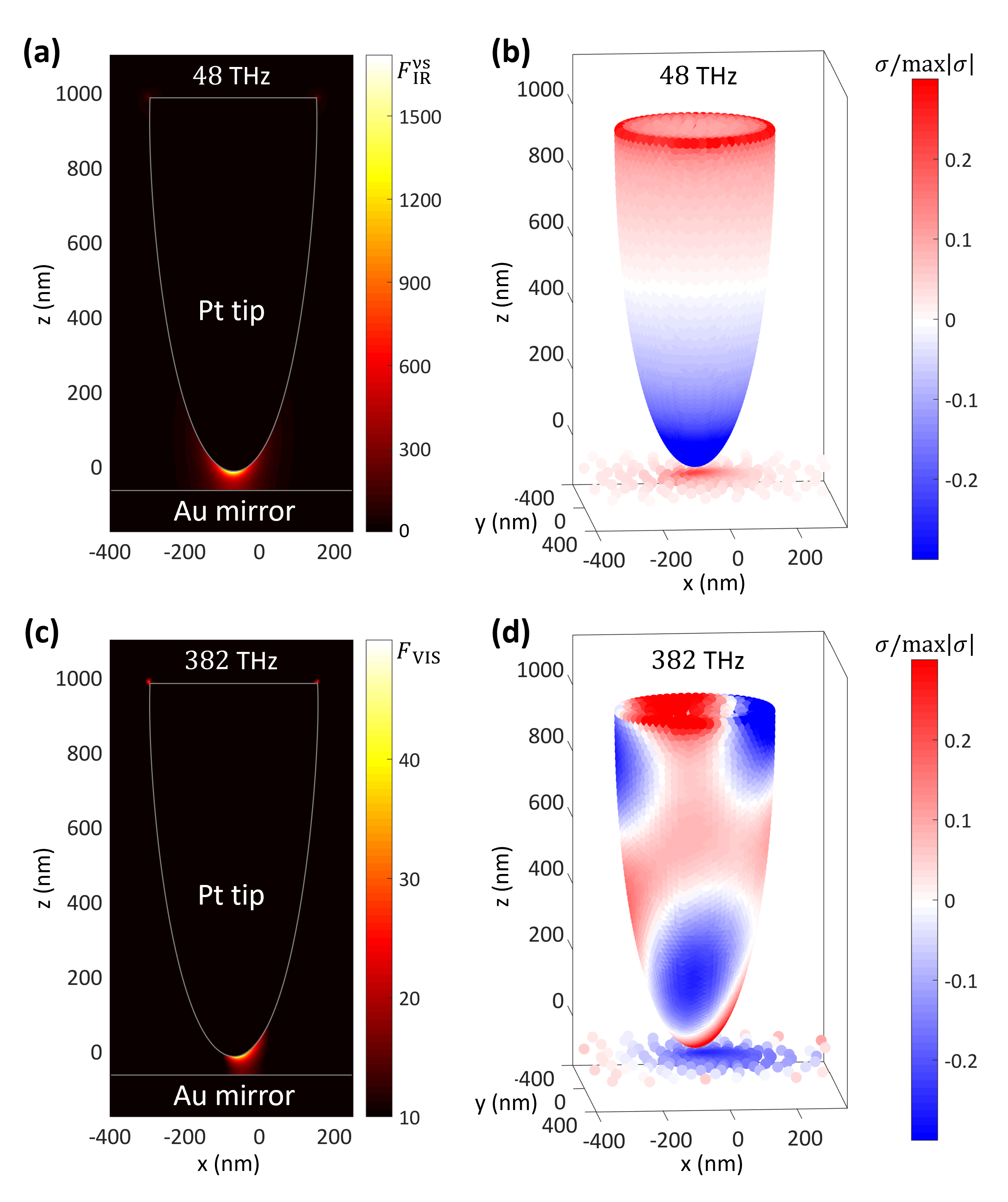}
  \caption{\textbf{Spatial distribution of intensity enhancement around the tip}. 
   Simulated spatial distribution of the intensity enhancement factors 
  $F_{\mathrm{IR}}^{\nu{\mathrm{s}}}$ (a) and $F_{\mathrm{VIS}}$ (c)
  and corresponding surface charges (b,d) around the Pt tip 
  in close proximity to a gold plane 
  ($\Delta z_{\mathrm{t}}=$48~nm) 
  under plane-wave illumination from the left side (35 degrees relative to sample surface) 
  at $\omega_{\mathrm{IR}}^{\nu\mathrm{s}}$ (a,b) and $\omega_{\mathrm{VIS}}$ (c,d).  
  We see that the location of maximum enhancement factor is below the tip apex for the IR calculation 
  but shifted in the direction of illumination for the VIS calculation. 
  This observation can be explained by the surface charge distributions of the tip, 
  which reveal that we have essentially a longitudinal (i.e. along tip axis) mode in IR 
  but both a longitudinal and a transversal mode in VIS. 
  The interference of the two modes leads to constructive interference on the right side of tip apex 
  and to destructive interference on left side of the tip apex.
  This finding explains why the tip has to be positioned on the left side of the nanoparticle 
  to provide maximum near-field illumination of the NPoM cavity in the VIS and to obtain maximum optical signals, 
  as observed in main text Fig.~2 (c,d) for $I^{\nu\mathrm{b}}_+$ signals.
  }
  \label{fig:sm-asymmetry}
\end{figure}


\newpage

\subsection{Tip-enhanced NPoM photoluminescence}

\begin{figure}[ht!]
\vspace*{-24pt}
\centering
\includegraphics[width=.85\textwidth]{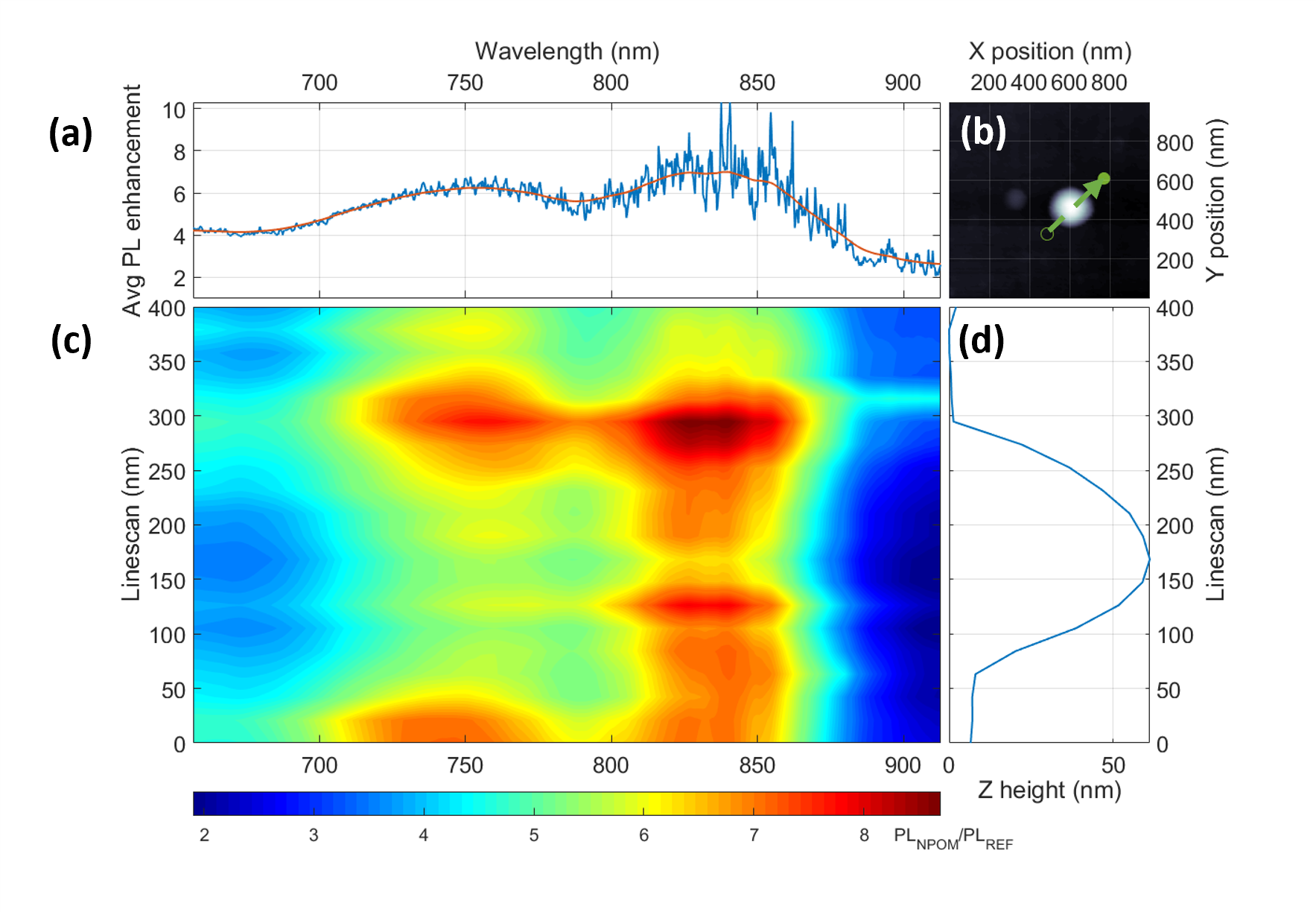}
\vspace*{-26pt}
  \caption{\textbf{Tip-enhanced NPoM photoluminescence}. 
In order to evidence the modification of the photonic local density of states (LDOS) 
of the NPoM cavity modes via the tip, we follow the method presented in Refs.~\cite{lumdee_gap-plasmon_2014,chen_intrinsic_2021_sm}. 
Under 561~nm laser illumination, interband transitions occur both in the NPoM cavity and in the bare gold film. 
The different NPoM modes enhance the gold photoluminescence of the gold film ($\mathrm{PL}_{\mathrm{REF}}$), 
i.e., the radiative recombination of the generated electron-hole pairs.  
and can be evidenced and characterized by the PL enhancement factor ($\mathrm{PL}_{\mathrm{NPOM}}/\mathrm{PL}_{\mathrm{REF}}$), 
as shown in this figure and in Fig.~\ref{fig:sm-pl-more}.  
  Fig.~\ref{fig:sm-pl} shows the spectral variations of $\mathrm{PL}_{\mathrm{NPoM}}/\mathrm{PL}_{\mathrm{REF}}$ 
  under illumination at 561~nm (50~$\mu$W) as a function of tip position for an isolated 80~nm gold nanosphere 
  on a BPT functionalized gold film. 
  (a) Averaged PL enhancement over the entire linescan 
  (raw signals in blue, PL results smoothed by a Whittaker filter in red). 
  (b) Topography image of the NPoM. The green arrow depicts the linescan trajectory. 
  (c) Waterfall plot of the PL enhancement as a function of tip position. 
  (d) Height profile acquired from the line scan. 
  Acquisition time per pixel is 50~s, $\mathrm{TA}=50$~nm. 
We attribute the two main features of the spectra around 750~nm and 840~nm 
to the L01 and S11 modes of the NPoM cavity, 
respectively \cite{tserkezis_hybridization_2015_sm}.  
Using these peak positions as well as the height of the NPoM 
as measured with the AFM (Fig.~\ref{fig:sm-pl}d) in the numerical calculations of the NPoM geometry, 
we can conclude that the optical gap size ($n\cdot d$) is about $1.61$~nm 
and the particle's facet size ($w$) is about $35$~nm.
We observe that the PL signals depend on the tip position, 
providing an additional evidence 
of the influence of the tip on the NPoM cavity modes. 
However, since the tip-enhanced PL process is different 
from the tip-enhanced processes studied in the main text, 
no direct comparison of the different enhancement factors can be made. 
  }
  \label{fig:sm-pl}
\end{figure}

\newpage

\begin{figure}[ht!]
\vspace*{-24pt}
\centering
\includegraphics[width=.85\textwidth]{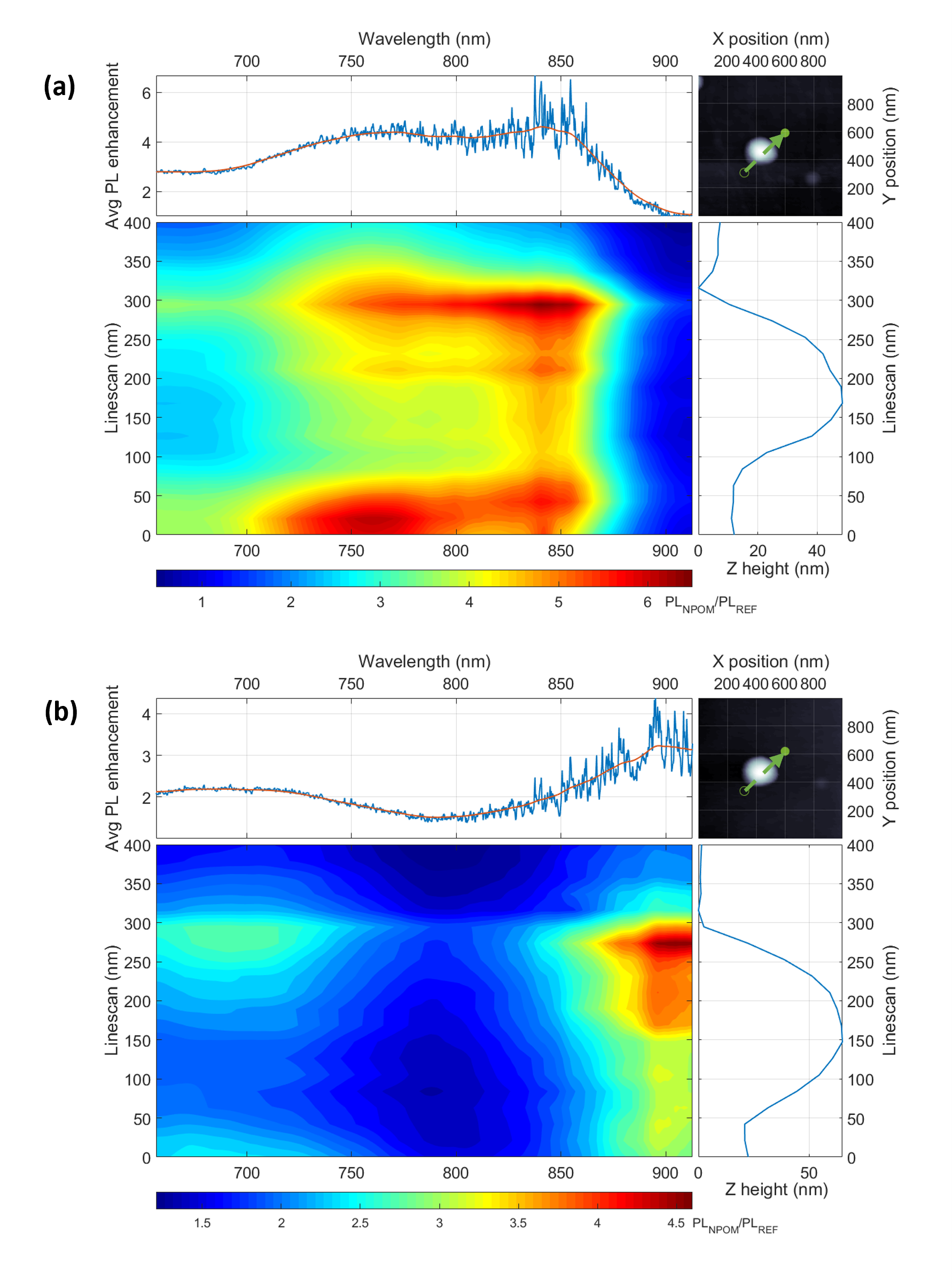}
\vspace*{-26pt}
  \caption{\textbf{Tip-enhanced photoluminescence (PL)}. 
  Spectral variations of $\mathrm{PL}_{\mathrm{NPoM}}/\mathrm{PL}_{\mathrm{REF}}$ 
  under illumination at 561~nm (50~$\mu$W) as a function of tip position 
  for an isolated gold nanosphere (a) 
  and an isolated gold nanocube (Nanopartz, 80~nm nominal side length) (b) 
  on a BPT functionalized gold film. 
  Panels and experimental parameters are described in Fig.~\ref{fig:sm-pl}. 
  }
  \label{fig:sm-pl-more}
\end{figure}

\newpage

\subsection{Scaling of intensity enhancement factors as a function of tip-NPoM distance}
\begin{figure}[ht!]
\centering
\includegraphics[scale=0.65]{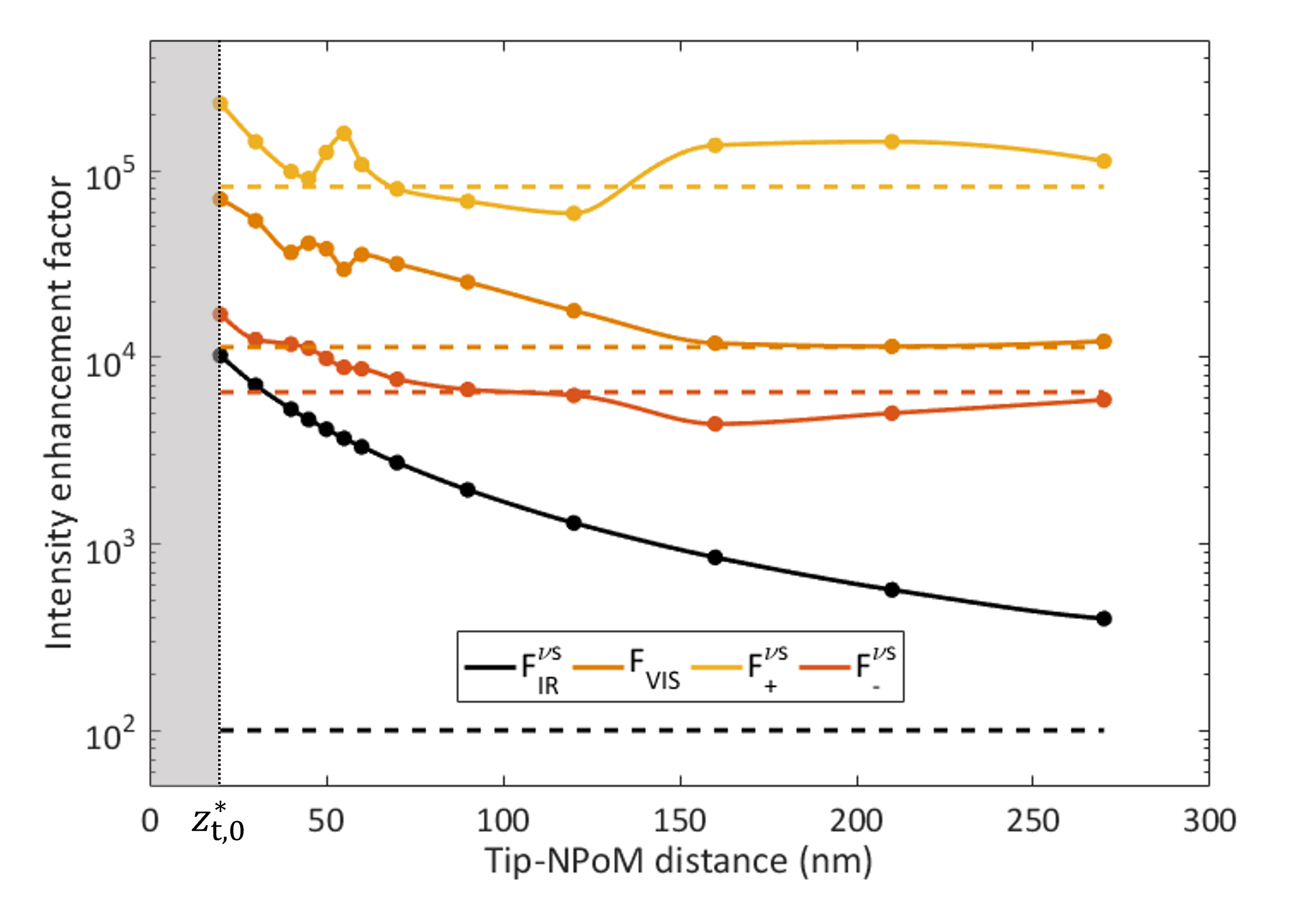}
  \caption{\textbf{Scaling of intensity enhancement factors as a function of tip-NPoM distance}. 
  Simulated intensity enhancement factors 
  $F_{\mathrm{IR}}^{\nu\mathrm{s}}(\vec{\mathrm{r}}_{\mathrm{hs}})$ (black curve), 
  $F_{\mathrm{VIS}}(\vec{\mathrm{r}}_{\mathrm{hs}})$ (orange),
  $F_{\mathrm{+}}^{\nu\mathrm{s}}(\vec{\mathrm{r}}_{\mathrm{hs}})$ (yellow),
  and $F_{\mathrm{-}}^{\nu\mathrm{s}}(\vec{\mathrm{r}}_{\mathrm{hs}})$ (red) 
  at the frequency of the vibrational mode $\nu_{\mathrm{s}}$ 
  as a function of vertical tip-NPoM distance. 
  The figure shows that, contrary to the monotonic decrease of the enhancement factor $F_{\mathrm{IR}}^{\nu\mathrm{s}}(\vec{\mathrm{r}}_{\mathrm{hs}})$
  for increasing tip-NPoM distance $\Delta z_{\mathrm{t}}$ in the IR spectral range, 
  the scaling behavior is complex in the visible spectral range 
  (here illustrated for $F_{\mathrm{VIS}}(\vec{\mathrm{r}}_{\mathrm{hs}})$,
  $F_{\mathrm{+}}^{\nu\mathrm{s}}(\vec{\mathrm{r}}_{\mathrm{hs}})$,
  and $F_{\mathrm{-}}^{\nu\mathrm{s}}(\vec{\mathrm{r}}_{\mathrm{hs}})$). 
  At large $\Delta z_{\mathrm{t}}$, the enhancement factors converge towards the NPoM cavity alone values.  
  Additional non-monotonic features are observed at lower distances 
  (for $40\ \mathrm{nm} \lesssim \Delta z_{\mathrm{t}} \lesssim 130\ \mathrm{nm}$) 
  and can be attributed to interferences between far-field illumination and tip illumination of the NPoM gap. 
  It is worth noting that even greater intensity enhancement factors are anticipated 
  for tip-NPoM distances smaller than $z^*_{\mathrm{t},0}\simeq20$~nm, the effective distance when operating our s-SNOM microscope in contact mode. This region, depicted by a grey area in the figure, remained out of reach for the current version of the microscope. 
  }
  \label{fig:sm-tbac-ef}
\end{figure}

%

\end{document}